\documentclass[12pt]{iopart}

\usepackage{epsfig} 
\usepackage{iopams}

\newcommand{\LAA}{\mathcal{L}_\mathrm{AA}}
\newcommand{\LFA}{\mathcal{L}_\mathrm{FA}}
\newcommand{\LbAA}{\tilde{\mathcal{L}}_\mathrm{AA}}
\newcommand{\LbFA}{\tilde{\mathcal{L}}_\mathrm{FA}}
\newcommand{\HbAA}{\tilde{H}_\mathrm{AA}}
\newcommand{\HbFA}{\tilde{H}_\mathrm{FA}}

\newcommand{\upket}{|\!\!\uparrow\rangle}
\newcommand{\downket}{|\!\!\downarrow\rangle}

\newcommand{\eq}{\mathrm{eq}}
\newcommand{\FA}{\mathrm{FA}} 
\renewcommand{\AA}{\mathrm{AA}} 

\newcommand{\bra}[1]{\langle #1 |} 
\newcommand{\ket}[1]{| #1 \rangle}

\newcommand{\bx}{\boldsymbol{x}} 
\newcommand{\by}{\boldsymbol{y}} 

\newcommand{\bk}{\boldsymbol{k}}

\renewcommand{\epsilon}{{\tilde{c}}}
\newcommand{\matx}[4]{\left(\begin{array}{cc}#4&#3\\#2&#1\end{array}\right)}
\newcommand{\vect}[2]{\left(\begin{array}{c}#2\\#1\end{array}\right)}

\newcommand{\sub}[1]{}
\newcommand{\bm}[1]{\mathbf{#1}} 
\newcommand{\iu}{\imath}

\begin{document}

\title[Mappings between reaction-diffusion and kinetically 
constrained systems]{Mappings between reaction-diffusion and 
kinetically constrained systems: $A+A\leftrightarrow A$ and 
the FA model have upper critical dimension $d_c=2$}

\author{Robert Jack$^1$, Peter Mayer$^2$ and Peter Sollich$^3$}

\address{$^1$ Rudolf Peierls Centre for Theoretical Physics, 
University of Oxford, 1 Keble Road, Oxford, OX1 3NP, UK}
\address{$^2$ Department of Chemistry, Columbia University, 3000
Broadway, New York, NY~10027, USA}
\address{$^3$ King's College London, Department of Mathematics,
London WC2R 2LS, UK}

\eads{\mailto{rljack@berkeley.edu}, \mailto{pm2214@columbia.edu}, \mailto{peter.sollich@kcl.ac.uk}}

\begin{abstract}
We present an exact mapping between two simple spin models:
the Fredrickson-Andersen (FA) model and a model of annihilating 
random walks with spontaneous creation from the vacuum, 
$A+A\leftrightarrow 0$. We
discuss the geometric structure of the mapping and its
consequences for symmetries of the models. Hence we are
able to show that 
the upper critical dimension of the FA model is two,
and that critical exponents 
are known exactly in all dimensions. These conclusions
also generalise to a mapping between $A+A\leftrightarrow 0$
and the reaction-diffusion system in which the
reactions are branching and coagulation, $A+A\leftrightarrow A$. 
We discuss the relation of our analysis to earlier work, and explain
why the models considered do not fall into the directed percolation
universality class.
\end{abstract}

\maketitle

The one spin facilitated
Fredrickson-Andersen (FA) model~\cite{FAModel,SollichReview} 
has been of interest recently as a simple 
model that exhibits dynamical 
heterogeneity~\cite{GCTheory,Whitelam04,Jung-SEB,FA-FDT,Toninelli-chi4}.
This model has a dynamical
critical point at zero temperature that is not characterised by
the divergence of any static lengthscale. As such, it represents
a possible
model of structural glasses, in which dynamical lengthscales seem
to be large, but no large static lengthscales have been 
found~\cite{GCTheory}.

An important advance was made by 
Whitelam~\emph{et~al.}~\cite{Whitelam04}. They showed that the master
equation of the FA model can be cast in a field-theoretic formalism that 
resembles a simple reaction-diffusion system with branching and 
coagulation proceses. The dynamical critical point occurs when the 
density of the diffusing defects vanishes. The properties of the system
near the fixed point can then be studied by the powerful methods of the 
renormalisation group. The analysis of Ref.~\cite{Whitelam04} 
indicated that fluctuation effects are important below four dimensions 
and that the model is controlled by the directed percolation (DP) fixed 
point~\cite{HinrichsenDP} between two and four dimensions.

In this article we follow Ref.~\cite{Whitelam04} in writing the FA 
model in a field theoretic language. The overall picture of a zero 
temperature dynamical fixed point that controls the low temperature 
scaling remains robust. However, we show that a somewhat hidden
symmetry of the FA 
model means that the fixed point governing the scaling is Gaussian 
above two dimensions, and identical to that of annihilation-diffusion below two 
dimensions. Hence the DP fixed point is not relevant 
to the FA model.

We determine the critical properties of the FA model
by means of an exact mapping to a
model of diffusing defects that appear in pairs from the vacuum,
and annihilate in pairs: we refer to this as
the AA (appear and annihilate) model. 
The mapping
holds in all dimensions and at all temperatures. 
In one dimension the AA model is a (classical) Ising chain
with particular single spin dynamics; above one dimension it is more
familiar as the reaction-diffusion system $A+A\leftrightarrow 0$. The critical
properties of the AA model were derived by Cardy and 
T\"{a}uber~\cite{Cardy98}; by using
our mapping we can then apply this derivation to the FA model.
The mapping also allows us to identify an (exact) duality symmetry
of the FA model, which it inherits from the parity symmetry of the AA model.
These symmetries are most simply expressed in terms of the master operators
for the stochastic processes, which can also be interpreted
as Hamiltonians for
quantum spin models~\cite{StinchcombeReview,Krebs95,Henkel97}.

We will show that the mapping from FA to AA models is a specific case
of a more general relationship between the reaction-diffusion 
processes in which the reactions are
$A+A\leftrightarrow 0$ on the one hand and $A+A\leftrightarrow A$ on
the other. (Note the
presence of both forward and reverse processes in these reactions.)
Our conclusion will be that these two systems are controlled by
the same fixed point, and that their critical exponents are therefore identical.

The form of the paper is as follows.
We define the FA and AA models in Section~\ref{sec:models}. Both are
models of hard core particles, but have generalisations
in which particles can share the same lattice site; we also
define these models (which we refer to as `bosonic').
The mapping between the FA and AA models is determined by their 
symmetries: 
it has quite a rich geometrical stucture, which we discuss in
Section~\ref{sec:map}. In Section~\ref{sec:fa_diff} it is shown that
the same mapping also connects the more general reaction-diffusion
processes $A+A\leftrightarrow A$ and  $A+A\leftrightarrow 0$.
An analagous mapping also exists between the corresponding bosonic
models (Section~\ref{sec:bos}). In Section~\ref{sec:crit}
we study the critical properties of the various models. We also
verify the scaling in three dimensions, where the 
differences between our predictions and those of 
Ref.~\cite{Whitelam04} are
clearest, by numerical simulation. Finally, we summarise our results
in Section~\ref{sec:conc}.

\section{The models}
\label{sec:models}

\subsection{Model definitions}

In this section we introduce several models that we will consider
in the remainder of the paper. We define them in terms of
microscopic dynamical rules, before outlining the methods
by which we represent 
their master equations and correlation functions.

We begin with the one-spin facilitated
Frederickson-Andersen model~\cite{FAModel}. This is referred to simply as
`the' FA model in what follows; more general FA models with
facilitation by several 
spins, which exhibit more cooperative behaviour~\cite{SollichReview}, will
not be covered here.
We define the model in terms of $N$ binary variables, $n_i\in \{0,1\}$,
on a hypercubic lattice in $d$ dimensions. The Hamiltonian for the system is 
trivial: 
\begin{equation} 
  E =\sum_i n_i. 
  \label{equ:E} 
\end{equation} 
We refer to a site with $n=1$ either
as an up-spin or as a `defect';
sites with $n=0$ are thought of as down spins or `empty'.
The spins can flip with Metropolis rates if and only if 
at least one of their neighbours is in the up state. That is,
for nearest neighbours $i$ and $j$, we have
\begin{equation}
  \begin{array}{cccl} 
    0_i 1_j & \rightarrow & 1_i 1_j, & \quad \mbox{rate } c, \\ 
    1_i 1_j & \rightarrow & 0_i 1_j, & \quad \mbox{rate } 1. 
  \end{array} 
  \label{FA_rates}
\end{equation}
Since we have detailed balance with respect to the Hamiltonian 
$E$, the
(dimensionless) rate $c$ parametrises the temperature according to
$c=\rme^{-\beta}$, with $\beta=1/T$ as usual. It also
sets the equilibrium density 
\begin{equation}
  \langle n_i \rangle_\mathrm{FA,eq} = 
  n_\eq \equiv \frac{c}{1+c}. 
  \label{equ:neq} 
\end{equation}
We have introduced the notation $\langle \cdot \rangle_\mathrm{FA,eq}$
for an equilibrium dynamical average.

To avoid confusion, we note that there are other versions of the FA model in
the literature. 
In the original model~\cite{FAModel} as described here
the rate for flipping a
spin $i$ is proportional to the number of its up-spin neighbours.
An alternative definition chooses the rate for flipping
spin $i$ to be independent of the number of neighbours in the up
state, as long as there is at least one such
neighbour~\cite{GraPicGra9397}. We do not  
believe that this choice makes any difference
to the critical behaviour of the system, but our exact mappings
apply only to the original model as defined above.

We can also define a variant of the FA model in which the occupations
$n_i$ are not restricted to binary values, but may be any non-negative
integers. We refer to this as the bosonic FA model, since the natural
field theory for describing it has bosonic fields. This model
has transition rates 
\begin{equation} 
  \begin{array}{cccl}
    n_i n_j & \rightarrow & (n_i+1)n_j, & \quad \mbox{rate } \epsilon \, n_j, \\
    (n_i+1) n_j & \rightarrow & n_i n_j, & \quad \mbox{rate } (n_i+1) n_j.
  \end{array} 
\label{bosonic_FA}
\end{equation}
Here and
throughout, parameters for the bosonic models are distinguished by
tildes from those for the hard core ones.
The bosonic model (\ref{bosonic_FA}) obeys 
detailed balance with respect to the
stationary state $P(\{n_i\})=\prod_i \rme^{-\epsilon} \, {\epsilon}^{n_i}/{n_i!}$, 
so the stationary state density is
\begin{equation}
\langle n_i \rangle_\mathrm{\widetilde{\FA},eq} = \epsilon.
\end{equation}
The bosonic stationary state is again a Gibbs distribution
with Hamiltonian $E$ and temperature defined by
$\epsilon=\rme^{-\beta}$. However, in contrast to the hard core case it
includes an a priori phase space weight factor $\prod_i 1/n_i!$, as
appropriate for boson statistics.

We will establish a mapping between the FA model and
another model of diffusing defects. In this model, defects appear in pairs
out of the empty state, and annihilate in pairs into it; they
also diffuse freely across the lattice. We refer to this model
as the AA model, since the defects appear and annihilate. Both the
AA and FA models can be interpreted as reaction-diffusion processes. 
The AA model has explicit diffusion, combined with reversible annihilation
$A+A\leftrightarrow 0$. (We follow the standard nomenclature of
reaction-diffusion models here, with $A$ referring to our single
species of particles, i.e.\ defects.) The model is defined for binary
variables: for nearest neighbours $i$ and $j$, 
\begin{equation}
  \begin{array}{cccl}
	  1_i 0_j & \rightarrow & 0_i 1_j, & \quad \mbox{rate } \gamma c', \\
		1_i 1_j & \rightarrow & 0_i 0_j, & \quad \mbox{rate } \gamma, \\
		0_i 0_j & \rightarrow & 1_i 1_j, & \quad \mbox{rate } \gamma c'^2, 
  \end{array} 
\end{equation}
where we choose 
\begin{equation}
c'=\frac{\sqrt{1+c}-1}{\sqrt{1+c}+1},\qquad 
\gamma=\frac{(1+\sqrt{1+c})^2}{2}=\frac{2}{(1-c')^2}.
\label{equ:def_gamma}
\end{equation}
This model also obeys detailed balance with respect
to the trivial Hamiltonian, $E$, at a temperature
parametrised by $c'$. Note that, once a trivial overall scale for
the rates has been removed, there are in principle two dimensionless
rates, and the requirement of detailed balance with respect to $E$
only fixes one of 
these, namely the ratio of the rates for appearance and
annihilation. For now, we do not let the diffusion rate
vary independently and instead tie $\gamma$ to $c'$. The mapping between
the FA and AA models will then connect 
models with the same value of the single parameter $c$.
The two-parameter generalisation of the AA model
with an arbitrary diffusion constant also has a mapping to
a generalised FA model, as we discuss in 
Section~\ref{sec:fa_diff}. For notational convenience
we study the above `standard' AA and FA models first in what follows.

Finally, there is also a bosonic variant of the AA model with rates 
\begin{equation} 
  \begin{array}{cccl} 
		n_i n_j & \rightarrow & (n_i-1) (n_j+1), & \quad \mbox{rate } \tilde{\gamma}  \epsilon' n_i, \\
		n_i n_j & \rightarrow & (n_i-1) (n_j-1), & \quad \mbox{rate } 
		       \tilde{\gamma} n_i n_j, \\
		n_i n_j & \rightarrow & (n_i+1) (n_j+1), & \quad \mbox{rate } \tilde{\gamma} \epsilon'^2,
  \end{array} 
\end{equation}
where we take $\epsilon'=\epsilon/4$ and $\tilde{\gamma}=2$. The
stationary state densities in the AA models are
\begin{equation}
\langle n_i \rangle_\mathrm{AA,eq} = n_\eq' \equiv
\frac{c'}{1+c'}
,\qquad
\langle n_i \rangle_\mathrm{\widetilde{\AA},eq} = 
\epsilon'.
\end{equation}
Note that in the limit of small $c$, the relation
(\ref{equ:def_gamma}) between $c'$ and $c$ becomes $c'=c/4$, and thus
identical to the one for the bosonic models. More generally, the hard
core and bosonic models should become effectively equivalent at low
densities where multiple occupany of sites is very unlikely. We will
exploit this correspondence frequently.

\subsection{Operator forms for the master operators}

It is convenient to write stochastic averages for systems such
as those defined above in an operator
formalism~\cite{Doi-Peliti}. This is a standard 
technique, so we largely restrict this section to definitions of the
quantities that we will use later. 

When considering the bosonic versions of the FA and AA models we use a
bosonic algebra with creation and annihilation operators on each site:
$[a_i,a_j]=[a_i^\dag,a_j^\dag]=0$; $[a_i,a_j^\dag]=\delta_{ij}$. The
state $\{n_i\}$ is then associated with the vector $\prod_i
(a_i^\dag)^{n_i}|0\rangle$, where $|0\rangle$ is the vacuum state
which has all sites empty; the set of all $2^N$ state probabilities $P(\{n_i\},t)$
is mapped to the vector $|\psi(t)\rangle =
\sum_{\{n_i\}} P(\{n_i\},t)\prod_i (a_i^\dag)^{n_i}|0\rangle$. The
individual probabilities can be retrieved via $P(\{n_i\},t)=\langle
0|\prod_i (a_i^{n_i}/n_i!)|\psi(t)\rangle$, and since they must sum to
unity one has $\bra{\tilde{e}}\psi(t)\rangle=1$ where
\begin{equation}
\bra{\tilde{e}} = \bra{0} \prod_i \rme^{a_i}
\end{equation}
is a `projection state' that implements the sum over all possible
system configurations. The master
equation can then be written in operator form as
$\partial_t|\psi(t)\rangle = -\mathcal{L}|\psi(t)\rangle$, where
$\mathcal{L}$ is known as the Liouvillian or simply the master
operator. The off-diagonal elements $-\langle 0|\prod_i
(a_i^{n'_i}/n'_i!)\mathcal{L}\prod_i (a_i^\dag)^{n_i}|0\rangle$ of
$-\mathcal{L}$ give the rates for transitions from state $\{n_i\}$ to
$\{n_i'\}$, while the diagonal elements follow from the
requirement $\bra{\tilde{e}} \mathcal{L}=0$. Since the
master equation is linear, it can be solved formally as $|\psi(t)\rangle =
\rme^{-\mathcal{L}t}|\psi(0)\rangle$. If we specify the initial state as
$\{n_i\}$ we can read off from this the probability of making a
transition to state $\{n'_i\}$ in some time interval $t$:
\begin{equation}
P_{\{n'_i\}\leftarrow\{n_i\}}(t) = \langle 0| \left[ \prod_i
\frac{a_i^{n_i'}}{n_i'!} \right] 
\rme^{-\mathcal{L}t} \left[ \prod_i (a_i^\dag)^{n_i} \right] |0\rangle.
\label{equ:PAA_bos}
\end{equation}
Expectation values over the stochastic dynamics can also be expressed in a
simple form; for example, the average of some function $f(\{n_i\})$ at
time $t$ becomes
\begin{equation}
\langle f(\{n_i\}) \rangle = \sum_{\{n_i\}} f(\{n_i\}) P(\{n_i\},t)
= \bra{\tilde{e}} f(\{\hat{n}_i\}) \rme^{-\mathcal{L}t}|\psi(0)\rangle,
\end{equation}
where $\hat{n}_i=a_i^\dag a_i$ is particle number operator for site
$i$. Similar expressions can be written for correlations
functions involving two or more times, as illustated below.

The hard core models with their binary occupation variables $n_i$ have
similar relations but here the states are generated by operators
$s^+_i$ and $s^-_i\equiv(s^+_i)^\dag$ in a spin-$\frac{1}{2}$ algebra, with
$(s^+_i)^2=(s^-_i)^2=0$ and $s^+_i s^-_i + s^-_i s^+_i=1$. The state
vector is now $|\psi(t)\rangle = \sum_{\{n_i\}} P(\{n_i\},t)\prod_i
(s^+_i)^{n_i}|0\rangle$, and conversely $P(\{n_i\},t) = \langle
0|\prod_i (s^-_i)^{n_i}|\psi(t)\rangle$. Conservation of probability
requires $\bra{e}\psi(t)\rangle =1$ with the appropriate projection
state now being
\begin{equation}
  \bra{e} = \bra{0} \prod_i (1+s^-_i).
  \label{equ:bra1} 
\end{equation}
Transition probabilities and single-time averages take the forms
\begin{equation}
P_{\{n'_i\}\leftarrow\{n_i\}}(t)
 = \langle 0 | \left[ \prod_i (s_i^-)^{n_i'} \right]
\rme^{-\mathcal{L}t} \left[ \prod_i (s_i^+)^{n_i} \right] | 0 \rangle,
\label{equ:PAA}
\end{equation}
and
\begin{equation}
\langle f(\{n_i\}) \rangle = \bra{e}
f(\{\hat{n}_i\}) \rme^{-\mathcal{L}t}|\psi(0)\rangle,
\end{equation}
respectively, where now $\hat{n}_i=s^+_i s^-_i$. Occasionally it will
be useful to write states and 
operators in notation analogous to Pauli matrices and spin vectors.
Choosing a basis at each site as 
\begin{equation}
\downket_i \equiv |0\rangle_i = \vect{0}{1}, 
\qquad \upket_i \equiv |1\rangle_i = \vect{1}{0},
\end{equation}
one has for example
\begin{equation}
s^+_i = \matx{0}{1}{0}{0}_i, \qquad
\hat{n}_i=\matx{1}{0}{0}{0}_i.
\end{equation}
In principle one should write in this expression
a direct product $\bigotimes_{j\neq i} I_j$ with identity operators at
all other sites, but for ease of presentation we drop this here
and below, along with site subscripts $i$ where these are clear from
the context. Our ordering of the basis states, while the reverse of the
usual convention for spins, facilitates comparisons with other work on
reaction-diffusion systems~\cite{Henkel97}. It also emphasises the
analogy to the bosonic case, where the only natural ordering of the
basis states is in order of increasing occupancy.

It remains to give the forms of the master operator $\mathcal{L}$ for
our models. Their matrix elements
are easily derived from the relevant transition rates as
explained above. One finds:
\begin{eqnarray}
\fl \LFA & = & \sum_{\langle ij \rangle} 
\left[ (s_i^+-1) s_i^- s_j^+ s_j^- s_i^+(s_i^--c)
+ (i\leftrightarrow j) \right], 
\label{LFA_def}
\\
\fl \LbFA &=& \sum_{\langle ij \rangle} 
\left[ (a_i^\dag-1) a_j^\dag a_j (a_i-\epsilon)
+ (i\leftrightarrow j) \right],
\label{LbFA_def}
\\
\fl \LAA &=& (\gamma/2) \sum_{\langle ij \rangle} 
\left[ (s_i^+-1)s^-_i(s_j^++1)s^-_j s^+_j(s_j^-+c')s^+_i(s_i^--c')
+ (i\leftrightarrow j) \right],
\label{LAA_def}
\\
\fl \LbAA &=& (\tilde\gamma/2) \sum_{\langle ij \rangle} 
\left[ (a_i^\dag-1) (a_j^\dag+1) (a_j+\epsilon') (a_i-\epsilon')
+ (i\leftrightarrow j) \right],
\label{LbAA_def}
\end{eqnarray}
where the sums run over all nearest neighbour pairs. 
The operators $\LFA$ and $\LAA$ for the hard core models have been written in 
a suggestive form that emphasises the connection with
their bosonic counterparts $\LbFA$ and $\LbAA$. 

Before leaving this section, we note that the stationary states of our
models have simple closed forms in the quantum formalism, viz.
\begin{equation}
  \ket{c} = \prod_i \frac{1+c \, s^+_i}{1+c} \ket{0} 
  \quad \mbox{and} \quad 
  \ket{\tilde{c}} = \prod_i \rme^{\tilde{c} (a_i^\dag - 1)} \ket{0}
  \label{equ:cket} 
\end{equation} 
for the hard core and bosonic case, respectively. The latter is
distinguished by a tilde as usual. Correlations in the
stationary state then also take a rather simple
form. For times $t_1,\ldots,t_k$ that are in increasing order we have
\begin{equation}
\fl \langle n_{i_1}(t_1) n_{i_2}(t_2) \dots n_{i_k}(t_k)\rangle_\mathrm{FA,eq} = 
\bra{e} \hat{n}_{i_k} \rme^{-\LFA(t_k-t_{k-1})} \hat{n}_{i_{k-1}} \cdots 
\hat{n}_{i_2} \rme^{-\LFA(t_2-t_1)} \hat{n}_{i_1} \ket{c}.
\label{equ:spin_correl}
\end{equation}
The AA model
has an identical relation with $\LFA$ replaced by
$\LAA$ and $c$ replaced by $c'$, and for the bosonic models one merely
has to substitute for the master operator, projection and stationary state
vectors their bosonic equivalents.

\section{Symmetries and mappings for hard core particles}
\label{sec:map}

\subsection{Detailed balance, parity and duality symmetries}

Having set up the operator formalism for dynamics, we now investigate
some properties of the Master operators for these models. We
first consider the effects of detailed balance, which tells us that
the operator $\mathcal{L}\rme^{-\beta\hat{E}}$ is
Hermitian (or more specifically symmetric, since all matrix elements are
real). Here $\hat{E}=\sum_i \hat{n}_i$ is the (Hermitian)
operator for the energy. Multiplying by $\rme^{\beta\hat{E}/2}$ from
the left and right shows that also 
\begin{equation}
H = \rme^{\beta \hat{E}/2} \mathcal{L} \rme^{-\beta\hat{E}/2} 
\end{equation}
is Hermitian. This is more useful than
$\mathcal{L}\rme^{-\beta\hat{E}}$ since it is related to the Liouvillian
$\mathcal{L}$ by a similarity transformation and
so has the same eigenvalues. For the FA model we can write explicitly
$\rme^{-\beta\hat{E}/2}=\prod_i h_i(c)$,
where $h_i(\cdot)$ is the single site operator
\begin{equation}
h_i(x) = x^{1/2} s^+_i s^-_i +  s^-_i s^+_i  
= \matx{x^{1/2}}{0}{0}{1}\sub{i}.
\label{equ:def_h}
\end{equation}
For the AA model we only need to replace $c$ by $c'$.  The
Hermitian forms of the Liouvillians are then
\begin{equation}
\fl H_\mathrm{FA} = \left[ \prod_i h_i^{-1}(c) \right]
\LFA \left[ \prod_i h_i(c) \right],
\qquad H_\mathrm{AA} = \left[ \prod_i h_i^{-1}(c') \right]
\LAA \left[ \prod_i h_i(c') \right].
\end{equation}
Their explicit forms make it evident that they are indeed
Hermitian: for example
\begin{equation}
H_\FA = 
\sum_{\langle ij \rangle} 
\left[ (s_i^+-\sqrt{c}) s_i^- s_j^+ s_j^- s_i^+(s_i^--\sqrt{c})
+ (i\leftrightarrow j) \right],
\label{H_FA}
\end{equation}
and $H_\AA$ is similarly obtained from $\LAA$ in (\ref{LAA_def}) by
replacing the coefficients $\pm 1$ and $\pm c'$ by $\pm \sqrt{c'}$.

The above similarity transformation to a Hermitian form of the Liouvillians
is convenient since it makes manifest the
symmetries and conserved quantites of the systems.
The mapping between FA and AA models relies on the fact that
the Hermitian operators $H_\mathrm{FA}$ and $H_\mathrm{AA}$
are related by the {\em exact} unitary (or, more specifically, orthogonal)
transformation
\begin{equation}
H_\mathrm{FA} = 
U^{-1}
H_\mathrm{AA} 
U,
\label{equ:def_u}
\end{equation}
where 
\begin{equation*}
U = \prod_i u_i , \quad
u_i = \frac{1}{\sqrt{1+c'}} \left[ 1 - 2 \iu \sqrt{c'}s^y_i \right] =  
\frac{1}{\sqrt{1+c'}} \matx{1}{-\sqrt{c'}}{\sqrt{c'}}{1}\sub{i},
\end{equation*}
with $s^y_i=(s^+_i-s^-_i)/2\iu$ as usual. 
Equation (\ref{equ:def_u}) is the key relation from which most other
results for the 
hard core models are derived; it is easy to verify by direct
calculation.
The operator $U$ has a simple geometrical interpretation: it is
just a rotation about the $y$-axis of the spin sphere, as illustrated in
Figure~\ref{fig:map_geom} below. 

From (\ref{equ:def_u}) we have directly a similarity transform between the
corresponding master operators for the FA and AA models:
\begin{equation}\LFA = V^{-1} \LAA V,
\label{equ:def_v}
\end{equation}
 with
\begin{equation*}
V = \prod_i v_i, \quad
v_i = \frac{\sqrt{1+c'}}{2}\sqrt{\frac{c}{c'}}\ h_i(c') u_i h_i^{-1}(c) =
\frac{1}{2}
\matx{1}{1-\sqrt{1+c}}{1}{1+\sqrt{1+c}}\sub{i}.
\end{equation*}
We have exploited the freedom to introduce an arbitrary prefactor into
$v_i$ to ensure that both its columns add up to unity, i.e.\
\begin{equation}
(\langle 0|_i+\langle 1|_i)v_i=\langle 0|_i+\langle 1|_i.
\label{v_stochastic}
\end{equation}
For the whole transformation $V$ this implies that
the projection state (\ref{equ:bra1}) is invariant under 
multiplication by either $V$ or $V^{-1}$ from the right, 
$\bra{e} V = \bra{e} V^{-1} = \bra{e}$. So 
(\ref{equ:def_v}) automatically maps a probability-preserving
Liouvillian onto another one. 

Various relations between correlation functions in the two models can
now be established. In addition to (\ref{v_stochastic}) one uses the
analogous property for application of $v_i$ to the steady state vector
on the right:
\begin{equation}
v_i \frac{|0\rangle_i+c|1\rangle_i}{1+c} = 
\frac{|0\rangle_i+c'|1\rangle_i}{1+c'}, 
\label{v_basis_effects}
\end{equation}
and hence $V \ket{c} = \ket{c'}$ for the steady state (\ref{equ:cket}). 
For the simplest connected
correlation function one has then, using the definition
(\ref{equ:spin_correl}) and the mapping (\ref{equ:def_v}),
\begin{eqnarray}
\fl \left\langle [n_{i}(t)-n_\eq] [n_{j}(0)-n_\eq]
\right\rangle_\mathrm{FA,eq} \nonumber \\ 
= \bra{e} (\hat{n}_i-n_\eq) \rme^{-\LFA t} (\hat{n}_j-n_\eq) \ket{c}, \nonumber \\
= \bra{e} V^{-1}[V(\hat{n}_i-n_\eq)V^{-1}] \rme^{-\LAA t} [V(\hat{n}_j-n_\eq)V^{-1}]
V \ket{c}.
\label{corr_fn_mapping_calc}
\end{eqnarray}
Equation~(\ref{v_stochastic}) implies that the leftmost factor of $V^{-1}$
can be absorbed into the projection state, while the rightmost factor $V$
just changes $c$ to $c'$ in the steady state vector.
Given that the number operators transform as
\begin{equation}
V\hat{n}_i V^{-1} = v_i \hat{n}_i v_i^{-1} =
\frac{1}{1+c'}\matx{1}{c'}{1}{c'}\sub{i},
\label{n_transf}
\end{equation}
one verifies also that
\begin{equation}
(\langle 0|_i+\langle 1|_i) v_i(\hat{n}_i-n_\eq)v_i^{-1} =
\frac{2}{\sqrt{1+c}}
(\langle 0|_i+\langle 1|_i) (\hat{n}_i-n'_\eq),
\label{appl_to_left}
\end{equation}
and
\begin{equation}
v_j(\hat{n}_j-n_\eq)v_j^{-1}\frac{|0\rangle_i+c'|1\rangle_i}{1+c'} =
\frac{2}{\sqrt{1+c}}
(\hat{n}_j-n'_\eq)\frac{|0\rangle_i+c'|1\rangle_i}{1+c'},
\label{appl_to_right}
\end{equation}
so that overall
\begin{equation}
\fl \left\langle [n_{i}(t)-n_\eq] [n_{j}(0)-n_\eq]
\right\rangle_\mathrm{FA,eq} = \frac{4}{1+c}
\left\langle [n_{i}(t)-n_\eq'] [n_{j}(0)-n_\eq'] \right\rangle_\mathrm{AA,eq}.
\label{equ:hc_prop_rel}
\end{equation}
It is then a trivial extension to show that for arbitrary connected
stationary state correlation functions of a single time difference,
\begin{eqnarray}
\fl \left\langle \Biggl[\prod_{r=1}^{l} (n_{i_r}(t)-n_\eq) \Biggr]
\Biggr[ \prod_{r=1}^m (n_{j_r}(0)-n_\eq) \Biggr]
\right\rangle_\mathrm{FA,eq} \nonumber \\ 
= 
\left(\frac{4}{1+c}\right)^{(l+m)/2}
\left\langle \Biggl[\prod_{r=1}^{l} (n_{i_r}(t)-n'_\eq) \Biggr]
\Biggr[ \prod_{r=1}^m (n_{j_r}(0)-n'_\eq) \Biggr]
\right\rangle_\mathrm{AA,eq}
\label{equ:map_correl}.
\end{eqnarray}
However, a direct generalisation to stationary state correlations involving
more than one time difference, or out-of-equilibrium quantities
depending on more than one time, is not possible. This is because the
transformation (\ref{n_transf}) of the number operator produces a
non-diagonal operator which does not directly correspond to a physical
observable. Only where the transformed operator is applied either to
the projection state on the left, as in (\ref{appl_to_left}), or the
steady state on the right, as in (\ref{appl_to_right}), can 
such a link be made; otherwise more complicated relations
result~\cite{MaySol06}.

The most useful aspect of the mapping (\ref{equ:def_v}) is that it
will enable us to reveal symmetries of the FA model which are
`inherited' from symmetries of the AA model. Specifically, 
it is clear from the dynamical rules of the AA model that the parity of
the total number of particles in the system is conserved. Mathematically,
we have that
\begin{equation}
\LAA = \left[\prod_i 2s^z_i \right] \LAA \left[\prod_i 2s^z_i \right],
\label{equ:parity}
\end{equation}
where $s^z_i=\hat{n}_i-\frac{1}{2}=\frac{1}{2} {{-1 \,\,\, 0} \choose {\hspace{1ex} 0 \,\,\, 1}}$. 
Geometrically, the operator $\prod_i 2s^z_i$ simply produces a rotation
of $\pi$ radians about the $z$-axis of the spin sphere. 

Since the FA and AA models are related by a similarity transformation,
there must be a symmetry of the FA model that is equivalent to
the AA parity symmetry.
Applying the transformation (\ref{equ:def_v}) to equation
(\ref{equ:parity}), we arrive at
\begin{equation} \LFA = W^{-1} \LFA W \label{equ:def_d}, \end{equation}
with
\begin{equation*}
\fl W = V^{-1} \left[\prod_i 2s_i^z\right] V = \prod_i w_i, \quad
w_i = v_i^{-1}(2s^z_i) v_i = \frac{1}{\sqrt{1+c}}
\matx{1}{-c}{-1}{-1}\sub{i}.
\end{equation*}
Note that $W^{-1}=W$, as expected for a symmetry deriving from
the parity symmetry in the AA model. To understand more closely the
effect of $W$ note first that, in the AA case, the rotation $\prod_i
2s^z_i$ maps the steady state vector $\ket{c'} \propto
\bigotimes_i(|0\rangle_i+c'|1\rangle_i)$ to the vector
$\bigotimes_i(|0\rangle_i-c'|1\rangle_i)$ where the probabilities of all
states $\{n_i\}$ containing an odd number of particles acquire
a negative sign. The sum and difference of these two states gives the
physical steady states for initial conditions containing an even and
odd number of particles, respectively. In the FA case, $W$ also maps
two steady states onto each other: $W\bigotimes_i(|0\rangle_i+c|1\rangle_i)
\propto |0\rangle = \bigotimes_i|0\rangle_i$. The symmetry thus links the
`conventional' steady state, which is reached for any nonzero initial
number of particles, to the vacuum, i.e.\ the empty state; the
latter is trivially a steady state since the kinetic constraints of
the FA model forbid any transitions into or out of it.
So while the original symmetry in the AA model connects steady states
that are basically equivalent, with associated `domains of attraction'
of equal size, the inherited symmetry of the FA model relates two very
different steady states, with one having a domain of attraction
containing {\em all} configurations except for the empty one.

\begin{figure}
\hfill \epsfig{file=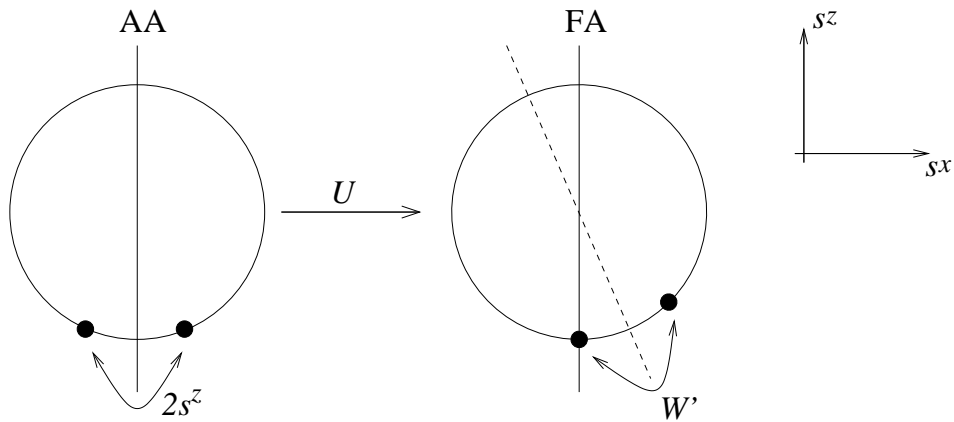,width=0.8\columnwidth}
\caption{The geometrical structure of the mapping
$U$ and the symmetries of the FA and AA models, in terms of
the Hermitian operators $H_\AA$ and $H_\FA$. Points on 
the spin sphere represent states $\rme^{-i\phi/2} \cos(\theta/2) \upket
+ \rme^{i\phi/2}\sin(\theta/2)\downket$ where $\theta$ and $\phi$
are the usual polar and azimuthal angles.
The black dots mark the position on the spin sphere of
the zero eigenstates of the operators; these are 
$\bigotimes_i (\downket_i \pm \sqrt{c'} \upket_i)$ for the AA model, and
$\bigotimes_i (\downket_i + \sqrt{c}\upket_i)$ and
$\bigotimes_i \downket_i$ for the FA model. Since these
states factorise over sites $i$, the figure can be read not just
schematically as representing the entire $N$-spin system, but also
literally as showing the spin spheres for a single site.
The rotation
$U$ is about the $y$-axis of the spin sphere (which points into the paper):
it maps $H_\AA$ onto $H_\FA$. The rotation $\prod_i (2s^z_i)$
of $\pi$ radians about the $z$-axis
maps $H_\AA$ onto itself. Applying the mapping $U$ gives the duality
transformation 
$W'=U^{-1}(\prod_i 2s_i^z)U$, which is a rotation of $\pi$ radians about
the dashed axis and maps $H_\FA$ onto itself. In
terms of the transformations in the main text, $W'$ is
simply the image $W$ after mapping $\LFA$ onto $H_\FA$:
we show $W'$ here since its geometrical structure is simpler. 
\label{fig:map_geom}
}
\end{figure}
The above relations between the FA and AA models, and their
corresponding symmetries, can also be understood in terms of the
associated Hermitian operators. They then have simple geometric
interpretations, as shown in Figure~\ref{fig:map_geom}.

In the following we will continue to refer to `the' steady state of
the FA model as the one with nonzero particle density. The symmetry
(\ref{equ:def_d}) then allows us to
relate the dynamics in this steady state to that in near empty
configurations. This implies relations between the associated
correlation functions. Proceeding as in (\ref{corr_fn_mapping_calc}),
one has for example
\begin{eqnarray}
\fl \left\langle [n_{i}(t)-n_\eq] [n_{j}(0)-n_\eq]
\right\rangle_\mathrm{FA,eq} \nonumber \\ 
= \bra{e} W^{-1} [W(\hat{n}_i-n_\eq)W^{-1}] \rme^{-\LFA t} [W(\hat{n}_j-n_\eq)W^{-1}] 
W \ket{c}, \nonumber \\
 = \bra{0} [W(\hat{n}_i-n_\eq)W^{-1}] \rme^{-\LFA t} [W(\hat{n}_j-n_\eq)W^{-1}] \ket{0}. 
\end{eqnarray}
Here we have used that application of $W$ to the steady state vector 
$\ket{c}$ on the right gives a multiple of the vacuum state. The same is easily 
checked for the operation of $W^{-1}$ on the projection state $\bra{e}$ on the
left; the associated proportionality factors cancel because of overall
normalisation. The transformation of the number operators is
\begin{equation*}
\fl W(\hat{n}_i-n_\eq)W^{-1} = w_i(\hat{n}_i-n_\eq)w_i^{-1} =
\frac{1}{1+c}\matx{1-c}{-c}{-1}{0} =
\frac{-cs^+_i-s^-_i+(1-c)\hat{n}_i}{1+c},
\end{equation*}
so that
\begin{equation}
\fl \left\langle [n_{i}(t)-n_\eq] [n_{j}(0)-n_\eq]
\right\rangle_\mathrm{FA,eq} =
\langle 0 | \left(-\frac{s^-_i}{1+c}\right) \rme^{-\LFA t}
\left(-\frac{c \, s^+_j}{1+c}\right) |0\rangle.
\label{equ:dual_prop}
\end{equation}
Up to the overall numerical factor $c/(1+c)^2$, the right hand side is of same
form as (\ref{equ:PAA}): it is the 
probability of a transition between particular initial and final
states, containing a single particle on sites $j$ and $i$
respectively. This relation generalises straightforwardly
to correlation functions involving more than two spatial points: we have that
\begin{eqnarray}
\fl \left\langle \left[\prod_{r=1}^{l} (n_{i_r}(t)-n_\eq) \right]
\left[ \prod_{r=1}^m (n_{j_r}(0)-n_\eq) \right]
\right\rangle_\mathrm{FA,eq} \nonumber \\ 
= \frac{(-1)^{l+m}c^m}{(1+c)^{l+m}}
\langle 0 | \left[\prod_{r=1}^l s^-_{i_r}\right] 
\rme^{-\LFA t}
\left[\prod_{r=1}^m s^+_{j_r}\right] |0\rangle, 
\label{equ:dual_correl}
\end{eqnarray}
where the right hand side is again of the form (\ref{equ:PAA}) and
gives the transition probability between an initial state with $m$
particles and a final state with $l$ particles. While
this relation may not be familiar, it is closely related to the
duality symmetry of the DP fixed point~\cite{HinrichsenDP}. The latter
is more usually expressed in terms of the dynamical action: see
Section~\ref{sec:crit}.

In summary, we see that the transformation $V$ maps the parity
symmetry of the AA model onto an (exact) duality symmetry of the FA
model. The mapping thus exposes a hidden symmetry which would not
easily be recognised by looking at the FA model alone.

\subsection{Models with additional diffusive processes}
\label{sec:fa_diff}

The discussion so far has considered the FA and AA models, both
defined in terms of a single parameter $c$. We now
generalise our arguments to models with extra diffusive processes.
This will show that our mapping applies more broadly
between reaction-diffusion models with, respectively, reversible
coagulation (i.e.\ coagulation and branching) and
reversible annihilation (i.e.\ annihilation and appearance) processes. The
generalised models will also allow us to elucidate the connection
between our mapping and related earlier studies.

Consider supplementing the FA model by an additional process
\begin{equation}
0_i 1_j \rightarrow 1_i 0_j, \qquad \hbox{rate } D.
\end{equation}
In particle language this is diffusion of a particle $A$ to a vacant
site, while the original processes (\ref{FA_rates}) are $A\to A+A$
(branching) and $A+A\to A$ (coagulation). This generalised FA model
can therefore also be viewed as 
the reaction-diffusion model $A+A\leftrightarrow A$.

The new diffusion term in the master operator can be written as
\begin{eqnarray}
\mathcal{L}_\mathrm{diff} &=& D \sum_{\langle ij\rangle}
(s^+_i - s^+_j) s^-_i s^-_j s^+_j s^+_i (s^-_i - s^-_j), \nonumber
\\
&=& -2D \sum_{\langle ij\rangle}
\left[ s^x_i s^x_j + s^y_i s^y_j + s^z_i s^z_j - 1/4 \right],
\label{Heisenberg}
\end{eqnarray}
where $s^x_i=(s^++s^-)/2$ and, as before, $s^y_i=(s^+_i-s^-_i)/2\iu$ and
$s^z_i=\hat{n}_i-1/2=(s_i^+s_i^- - s_i^-s_i^+)/2$.
We recognise in (\ref{Heisenberg}) the Heisenberg model: see Ref.~\cite{StinchcombeReview}
for a summary of the links between the properties of stochastic systems
and their corresponding quantum spin Hamiltonians.

For our purposes it is important to recognise that $\mathcal{L}_\mathrm{diff}$
has nonzero matrix elements only between states containing
equal numbers of particles; it is therefore invariant under transformation with
$\prod_i h_i(x)$ -- so that the associated Hermitian operator
is identical to $\mathcal{L}_\mathrm{diff}$ -- and under the parity
transformation $\prod_i 2s^z_i$. Due to its Heisenberg form, $\mathcal{L}_\mathrm{diff}$ is also
left invariant by any  
global spin rotation, and in particular by $U$. Combining these
properties, invariance under $V$ and $W$ then also follow.  
Hence the structure
of the preceding subsection is all 
preserved for the generalised models: the 
generalised FA model with diffusion rate $D$, branching rate $c$ and
coagulation rate $1$ maps via $V$ onto a generalised AA model with rates
\begin{equation}
  \begin{array}{cccl}
		1_i 0_j & \rightarrow & 0_i 1_j, & \quad \hbox{rate } \gamma c' + D, \\
		1_i 1_j & \rightarrow & 0_i 0_j, & \quad \hbox{rate } \gamma, \\
		0_i 0_j & \rightarrow & 1_i 1_j, & \quad \hbox{rate } \gamma c'^2, 
  \end{array} 
\end{equation}
where $\gamma$ and $c'$ depend only on $c$, as defined in
(\ref{equ:def_gamma}). 
We note that all generalised FA models have conjugate AA models,
but that AA models in which the rate for the diffusive process is
less than $\gamma c'$ cannot be mapped to FA models with positive rates. 

At this point, we make contact with two earlier studies.
Krebs~\emph{et al.}~\cite{Krebs95} studied the above generalised
models at zero temperature but with nonzero $D$. Consistent with this,
their mapping between the models is the limit of our mapping $V$ for
$c,c'\to 0$. Henkel~\emph{et~al.}~\cite{Henkel97} 
implicitly had the full mapping $V$, but considered it only in the
context of one-dimensional systems that are solvable by free fermions.
The AA model then reduces to the Glauber-Ising chain (their model IV) and
the relevant generalised FA model has $D=1$
(their model II). Henkel~\emph{et~al.} did not comment that
the mapping applies to all dimensions and to arbitrary values of the
diffusion constant.

\begin{figure}
\hfill \epsfig{file=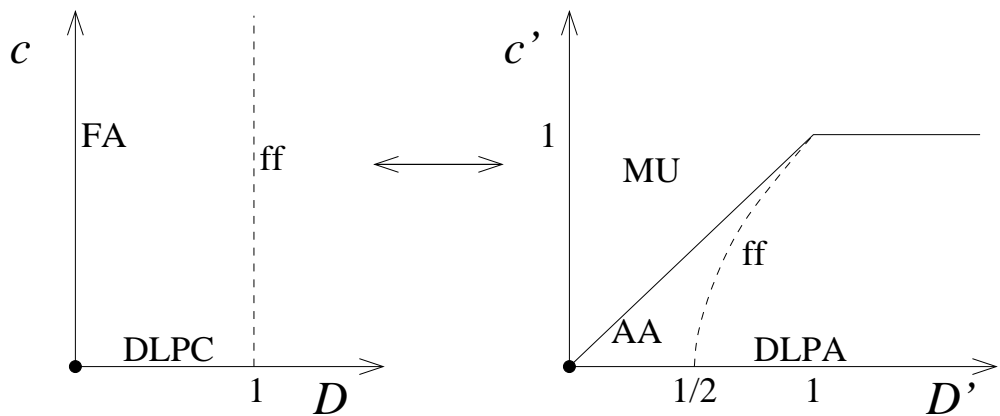,width=0.84\columnwidth}
\caption{Sketch of the mapping between the two-parameter
families of generalised FA ($A+A\leftrightarrow A$) and AA
($A+A\leftrightarrow 0$) models. Generalised FA models
(with non-negative $c$) map onto generalised AA models with $0\leq c'<1$
and $c'\leq D'$. The standard FA models with $D=0$ 
map to models on the standard AA line $c'=D'$. 
The lines $c=0$ and $c'=0$ correspond to 
diffusion-limited pair annihilation (DLPA) and
diffusion-limited pair coagulation (DLPC) respectively~\cite{Krebs95}. 
In one dimension the models are solvable by free fermions
on the lines marked `ff'. These lines are given by $D=1$ and $D'=(1+c'^2)/2$, and
the mapping transforms them into each other~\cite{Henkel97}. In the region
marked MU the mapping is unphysical: such generalised AA models
do not have FA counterparts with positive rates, though it seems
unlikely that this would have physical consequences for the behaviour
of the corresponding AA model. 
}
\label{fig:Dc_plane}
\end{figure}

We illustrate 
the relation between our work and Refs.~\cite{Krebs95}
and~\cite{Henkel97}
in Figure~\ref{fig:Dc_plane}. We parametrise the FA
($A+A\leftrightarrow A$)
models by the ratio of branching and coagulation rates $c$ and
the ratio of diffusion and coagulation rates $D$. In the AA
case, appropriate dimensionless parameters are the ratio of appearance
and annihilation rates 
$c'^2$ and the ratio of diffusion and annhilation rates
$D'=c'+D/\gamma$. All generalised
FA models map onto generalised AA models with
$c'<1$ and $D'\geq c'$. The standard FA
model is $D=0$, giving $D'=c'$, while the pure coagulation/annihilation
models of Ref.~\cite{Krebs95} correspond to $c=c'=0$. The free fermion
condition of Ref.~\cite{Henkel97}, finally, is the line $D=1$
which maps onto the Glauber-Ising line $D'=(1+c'^2)/2$. (The Glauber-Ising
chain has diffusion rate $1/2$ and annihilation rate
$(1+c'^2)^{-1}$, giving the stated ratio.)

To see the explicit link between our mapping and that of
Ref.~\cite{Henkel97}, one notes that their free-fermion quantum
Hamiltonian is directly in the form of the Liouvillian $\LAA$ for the
AA model if its parameters are chosen as $D_1=D_2=1$,
$h_1=h_2=(1-c'^2)/(1+c'^2)$, $\eta_1=2/(1+c'^2)$ and $\eta_2=\eta_1
c'^2$. Henkel {\em et al.}~\cite{Henkel97} then show that the FA Liouvillian can be
obtained by the similarity transformation $\LFA=B\LAA B^{-1}$, with
$B=\prod_i b_i$ and
\begin{equation}
b_i = \matx{1/\sqrt{a}}{0}{0}{\sqrt{a}}
\matx{\cosh\phi}{\sinh\phi}{\sinh\phi}{\cosh\phi}
\matx{1/b}{0}{0}{b}.
\end{equation}
Following through their analysis gives $a=\rme^{\iu\pi/2}c^{-1/2}$,
$b=\rme^{-\iu\pi/4}c'^{1/4}$ and $\phi=\iu\theta$, where $\tan
2\theta=\sqrt{c}$. The latter condition can also be written as
$\tan\theta=\sqrt{c'}$, so that $\theta$ is in fact the rotation angle
associated with our mapping $U$. Inserting these values one has
$b_i\propto v_i^{-1}$ (with proportionality factor
$[(1+c)/4]^{1/4}$) as expected by comparison with our Equation~(\ref{equ:def_v}).
We note that Henkel {\em et al.} associate particles with {\em down}
spins rather than up spins as here. However, because they also use the
opposite (i.e., conventional) ordering of the two local basis vectors
$\upket$ and $\downket$, the matrix
representations of all operators are the same.

To summarise, we showed in this section that the FA and AA models have
quite a rich  
geometric structure underlying their symmetries and the
relations between them. These relations further extend to a general mapping
between reaction-diffusion models with coagulation and branching
($A+A\leftrightarrow A$) and annihilation and appearance
($A+A\leftrightarrow 0$).
We expect the critical behaviour (at small particle densities, i.e.\
low temperature)
of these models to be determined by their
symmetry properties. However, the hard core constraint that allows only one
particle per site makes an explicit renormalisation group analysis of
such critical properties awkward. We therefore show next that the bosonic
models, where this constraint is removed, have analogous symmetries
and mappings between coagulation and annihilation models.

\section{Bosonic models}
\label{sec:bos}
The bosonic models introduced in Section~\ref{sec:models} 
have similar properties 
to those discussed for hard core (spin) models in the previous
section. The Liouvillians again have Hermitian analogues defined by 
$H = \rme^{\beta \hat{E}/2} \mathcal{L} \rme^{-\beta\hat{E}/2}$. The
energy operator is now
$\hat{E}=\sum_i a_i^\dag a_i$, so if we define the bosonic version of
$h_i(x)$ as 
\begin{equation}
\tilde{h}_i(x) = x^{a_i^\dag a_i/2},
\label{equ:def_ht}
\end{equation}
then
\begin{equation}
\fl \HbFA = \left[\prod_i \tilde{h}_i^{-1}(\epsilon)\right] \LbFA
\left[\prod_i \tilde{h}_i(\epsilon)\right], \qquad
\HbAA = \left[\prod_i \tilde{h}_i^{-1}(\epsilon')\right] \LbFA
\left[\prod_i \tilde{h}_i(\epsilon')\right].
\end{equation}
$\HbFA$ has the same form as (\ref{LbFA_def})
except that the numerical constants 1 and $\epsilon$ are
replaced by $\sqrt{\epsilon}$, i.e.
\begin{equation}
\HbFA = \sum_{\langle ij\rangle} [ 
( a^\dag_i - \sqrt{\epsilon} ) a^\dag_j a_j (a_i - \sqrt{\epsilon})
+(i\leftrightarrow j) ],
\label{HbFA}
\end{equation}
and $\HbAA$ is obtained similarly from
(\ref{LbAA_def}). This is analogous to the hard core case, but easier
to see for the bosonic models since the transformation by $h_i(x)$ simply
rescales particle creation and annihilation operators, according to
the first of the relations
\begin{eqnarray}
\rme^{\lambda a^\dag a} F(a,a^\dag) \rme^{-\lambda a^\dag a} &=&
F(\rme^{-\lambda}a,\rme^{\lambda} a^\dag),
\label{n_exp}
\\
\rme^{\lambda a + \mu a^\dag} F(a,a^\dag) \rme^{-\lambda a - \mu a^\dag} &=&
F(a-\mu,a^\dag+\lambda).
\label{a_adag_exp}
\end{eqnarray}
One expects that at low particle densities
$n_\eq=c/(1+c)\approx c$ the constraint of at most single occupancy in
the hard core FA model will be irrelevant, so that it becomes
equivalent to the corresponding bosonic model with $\epsilon=c$; the same
argument applies to the hard core and bosonic AA models. This
physical reasoning~\cite{Whitelam04} can be further supported by a
large-$S$ expansion of the hard core models (\ref{sec:large_S}).

We now discuss the mappings between the bosonic models and their symmetries.
The main conclusion is that the structure of the hard core models is
preserved in their bosonic counterparts. 
The basic mapping between the two bosonic Hermitian operators is 
\begin{equation}
\HbFA = \tilde{U}^{-1} \HbAA \tilde{U},
\end{equation}
with the unitary operator
\begin{equation}
\tilde{U}=\prod_i \tilde{u}_i, \qquad
\tilde{u}_i = \rme^{(a_i-a^\dag_i)\sqrt{\epsilon}/2}.
\end{equation}
This is easy to verify, bearing in mind that $\epsilon'=\epsilon/4$
and $\tilde{\gamma}=2$: from (\ref{a_adag_exp}), the transformation by
$\tilde{U}$ shifts all $a_i$ and $a_i^\dag$ in $\HbAA$ by
$-\sqrt{\epsilon}/2=-\sqrt{\epsilon'}$.
If one uses a basis of bosonic coherent states then the mapping
is a translation in the complex plane that parametrises these states.
This is the analogue of the rotation of the spin sphere generated
by $u_i$, consistent with the intuition (\ref{sec:large_S}) that
the bosonic models effectively `flatten' the spin sphere onto the
complex plane of coherent states. 

Combining $\tilde{u}_i$ and $\tilde{h}_i$ we have a
relation between the Liouvillians,
\begin{equation}
\fl \LbFA = \tilde{V}^{-1} \LbAA \tilde{V}, \qquad \tilde{V} = 
\prod_i \tilde{v}_i, \qquad 
\tilde{v}_i=\tilde{h}_i(\epsilon') \tilde{u}_i \tilde{h}^{-1}_i(\epsilon) 
= 2^{-a^\dag_i a_i} \rme^{(a_i-\epsilon a^\dag_i)/2}, 
\label{equ:def_vt}
\end{equation}
where the explicit form of $v_i$ follows using (\ref{n_exp}) and
(\ref{a_adag_exp}). These relations also show that the transformation
by $V$ is simply a combined 
shift and rescaling of the bosonic operators
\begin{equation}
\tilde{V}^{-1} a_i \tilde{V} = \frac{1}{2}\left(a_i -
\frac{\epsilon}{2}\right), \qquad \tilde{V}^{-1} a_i^\dag \tilde{V} =  
2\left(a_i^\dag - \frac{1}{2}\right).
\label{equ:bos_op_map}
\end{equation}
The mapping between the two Liouvillians again relates the
parity symmetry of the bosonic AA model,
\begin{equation}
\LbAA = (-1)^{\sum_i a^\dag_i a_i} \LbAA (-1)^{\sum_i a^\dag_i a_i},
\end{equation}
to the duality symmetry of the bosonic FA model
\begin{equation}
\LbFA = \tilde{W}^{-1} \LbFA \tilde{W}, \qquad
\tilde{W} = \tilde{V}^{-1} (-1)^{\sum_i a^\dag_i a_i} \tilde{V}.
\label{equ:def_dt}
\end{equation}
One finds explicitly
\begin{equation}
\tilde{W} = \tilde{W}^{-1} = \prod_i (-1)^{a^\dag_i a_i}
\rme^{a_i-\epsilon a_i^\dag},
\end{equation}
and the duality symmetry transforms the bosonic operators as
\begin{equation}
\tilde{W}^{-1} a_i \tilde{W} = \epsilon-a_i, \qquad
\tilde{W}^{-1} a_i^\dag \tilde{W} = 1-a_i^\dag.
\label{equ:bos_op_duality_map}
\end{equation}

The mappings $\tilde{V}$ and $\tilde{W}$ allow us to establish the
analogues of (\ref{equ:map_correl}) and (\ref{equ:dual_correl}). The
first of these relates steady state correlations in the bosonic FA and
AA models via
\begin{eqnarray}
\fl \left\langle \Biggl[\prod_{r=1}^{l} (n_{i_r}(t)-\epsilon) \Biggr]
\Biggr[ \prod_{r=1}^m (n_{j_r}(0)-\epsilon) \Biggr]
\right\rangle_{\widetilde{\FA},\eq} \nonumber \\ 
= 
2^{l+m} \left\langle \Biggl[\prod_{r=1}^{l} (n_{i_r}(t)-\epsilon') \Biggr]
\Biggr[ \prod_{r=1}^m (n_{j_r}(0)-\epsilon') \Biggr]
\right\rangle_{\widetilde{\AA},\eq}.
\label{equ:map_correl_bos}
\end{eqnarray}
Here the $i_r$ must label sites that are all distinct from each other,
as do the $j_r$, though the two sets may contain sites in common with each
other. 
The prefactor on the right agrees with the low density limit of
the one in (\ref{equ:map_correl}), supporting our intuition about
the equivalence of hard core and bosonic models in this regime.

The duality symmetry of the bosonic FA model results in
\begin{eqnarray}
\fl \left\langle \Biggl[\prod_{r=1}^{l} (n_{i_r}(t)-\epsilon) \Biggr]
\Biggr[ \prod_{r=1}^m (n_{j_r}(0)-\epsilon) \Biggr]
\right\rangle_{\widetilde{\FA},\eq} \nonumber \\ 
=(-1)^{l+m}\epsilon^m \langle 0| \left[\prod_{r=1}^l a_{i_r}\right] 
\rme^{-\LbFA t}
\left[\prod_{r=1}^m a^\dag_{j_r}\right] |0\rangle. 
\label{equ:dual_correl_bos}
\end{eqnarray}
This again relates steady state correlations to transition probabilities
between specific initial and final states; the prefactor approaches
the one in (\ref{equ:dual_correl}) for $c=\epsilon\to 0$.

Following our discussion in Section~\ref{sec:fa_diff}, one expects that
the structure of the bosonic mapping will be preserved if an extra diffusive
process is added to both models. This is easily verified. It
is also immediate to show that the mapping is unchanged if 
we add on-site branching
and coagulation processes, as long as we retain detailed balance for the
whole model. These processes then map to on-site appearance and annihilation
in the generalised AA model. (Recall that our standard bosonic FA and AA models
have processes that always act on pairs of sites.)

The similarities between Eqs (\ref{equ:def_ht}-\ref{equ:dual_correl_bos}) and 
(\ref{equ:def_h}-\ref{equ:def_d}) 
show clearly that the bosonic models have the same structure
as those with hard core exclusion. In the next section we consider
the critical properties of these bosonic models; from our arguments 
the critical properties of the hard core models should be identical,
and we check this by comparing our predictions to numerical simulation.

\section{Critical properties}
\label{sec:crit}

We have shown that the FA and AA models are linked by an exact
mapping. Now, both models have scaling behaviour at small defect densities that
is characterised by the fixed point of a renormalisation group (RG) flow.
In Section~\ref{sec:rg} we use the mapping of the previous section together
with known results to show that the FA model has
upper critical dimension
$d_c=2$; this conclusion also applies to the generalised FA model, i.e.\
the reaction-diffusion model $A+A\leftrightarrow A$. The critical
scaling is then characterised by 
the well-known mean-field (Gaussian) exponents in $d>2$;
we also derive exact exponents below $d_c$ that coincide with
known results in one dimension. 
Our results differ from earlier studies in two and three dimensions:
in Section~\ref{sec:gauss} we therefore use simulations to confirm the
predicted mean-field scaling in $d=3$. 
In Section~\ref{sec:scaling}
we derive some analytical results for the scaling limit of 
correlation functions in $d>2$. 
Finally we discuss in Section~\ref{sec:pers} the scaling of the persistence 
function since data for this were used in Ref.~\cite{Whitelam04} to
support the argument that non-mean-field fluctuation corrections are
significant in three dimensions.

\subsection{Renormalisation group analysis}
\label{sec:rg}

The critical properties of the bosonic AA model were established by Cardy
and T\"{a}uber~\cite{Cardy98}: 
in their notation the model corresponds to $k=2$, $\tau>0$, $\sigma_m=0$. 
We write the generating functional for dynamical correlations in the
stationary state as a path integral on the lattice
\begin{equation}
Z_\AA = \int \mathcal{D}[\{\varphi_{it}\},\{\varphi^\dag_{it}\}]\,
\rme^{-\int\rmd t\, \left\{
\left[\sum_i \varphi^\dag_{it} \partial_t \varphi_{it} \right]
+ L_\AA[\{\varphi_{it}\},\{\varphi_{it}\}] \right\} }
\end{equation}
where $\varphi_{it}$ and $\varphi_{it}^\dag$ are time-dependent
conjugate fields at each
site $i$. The `Lagrangian'
$L_\AA[\{\varphi_{it}\},\{\varphi_{it}^\dag\}]$ is obtained from
$\LAA$ by replacing $a_i\to \varphi_{it}$ and $a_i^\dag\to
\varphi_{it}^\dag$. 
and depends on all the fields at a single time.

Taking the continuum limit, the lattice fields $\{\varphi_{it}\}$ are
promoted to a field $\phi_{xt}$ depending on spatial position $x$ and time $t$.
The generating functional becomes
\begin{equation}
Z_\AA =  \int \mathcal{D}[\phi_{xt},\phi^\dag_{xt}] \,\rme^{-S_\AA[\phi_{xt},\phi^\dag_{xt}]}
\end{equation}
where the functional $S_\AA$ is known as the dynamical action.
Including gradient terms up to second order gives
\begin{eqnarray}
\fl S_\AA[\phi,\phi^\dag] = \int\rmd^dx\,\rmd t\, \phi_{xt}^\dag \partial_t \phi_{xt}
\nonumber\\
+ \lambda_0 (\phi^\dag_{xt} - 1)(\phi_{xt}-\rho')(1+l_0^2\nabla^2/2) (\phi^\dag_{xt} + 1)(\phi_{xt}+\rho')
\label{equ:saa_bare}
\end{eqnarray}
where we have neglected boundary terms;
$\rho'$ is the steady state density (proportional
to $\epsilon'$), $l_0$ is
the microscopic lengthscale (lattice spacing) and 
$\lambda_0$ is a bare
coupling constant that sets the microscopic timescale.
The dimensions of $\lambda_0$ are [time]$^{-1}$[length]$^d$;
the field $\phi^\dag$ is chosen to be
dimensionless and $\phi$ has dimension of
[length]$^{-d}$. 

While the above factorised form for the action was useful for the
exact mappings of the previous sections, the RG calculation requires
us to separate the terms in the action that correspond to different physical
processes. We write
\begin{eqnarray}
\fl S_\AA[\phi,\phi^\dag] = \int\rmd^dx\,\rmd t\, \left\{
\phi_{xt}^\dag (\partial_t
-\lambda_0 \rho' l_0^2 \nabla^2) \phi_{xt}
+ \lambda_0 [(\phi^\dag_{xt})^2 - 1](\phi_{xt}^2-\rho'^2)
+ L_{\AA,1} \right\}
\label{equ:saa}
\end{eqnarray}
where
\begin{equation}
L_{\AA,1} = (\lambda_0 l_0^2/2)[ \phi^\dag_{xt}\phi_{xt}\nabla^2\phi^\dag_{xt}\phi_{xt}
+(\nabla\phi_{xt})^2+(\rho'\nabla\phi^\dag_{xt})^2]
\end{equation}
(we continue to ignore boundary terms when integrating by parts over spatial
degrees of freedom).
Physically,
we recognise the first term in (\ref{equ:saa}) as a diffusive propagator
for the excitations and the second term as local appearance and annihilation
processes. The terms contained in $L_{\AA,1}$
will be irrelevant in the RG sense since their only effect
is to enforce the fact that appearance
and annihilation of excitations take place on pairs of adjacent sites and not
on single sites. 
(In terms of the RG calculation these terms modify the 
spatial structure of terms in the action that are already present, but they
are not responsible for new terms, or for any singular behaviour.)
We therefore neglect $L_{\AA,1}$ and
arrive at the action considered in Ref.~\cite{Cardy98} for the case
($k=2$, $\tau>0$, $\sigma_m=0$) described above.

We now follow Ref.~\cite{Cardy98} in renormalising this dynamical action.
The parity symmetry of the bosonic AA model is 
\begin{equation}
S_\AA[\phi_{xt},\phi^\dag_{xt}] = S_\AA[-\phi_{xt},-\phi^\dag_{xt}],
\end{equation}
and must not
be obscured by making any shift of the fields~\cite{Cardy98}. The symmetry
is clearly preserved under the RG flow so only terms in the action
with this symmetry need be considered. Power counting then
shows that the upper critical dimension will be two. Above
$d=2$, therefore, the critical exponents have their
mean-field values
\begin{equation}
(z,\nu,\beta)_{d>2}=(2,1/2,1).
\end{equation}
Here we have defined $\beta$ 
by the scaling of the steady-state density $\lim_{t\to\infty}\langle
n_i(t) \rangle \sim \rho'^\beta$,
and $\nu$ by the correlation length scaling $\xi \sim \rho'^{-\nu}$. 
(The notation
$a\sim b$ means that $a$ is proportional to $b$ in the scaling
limit, i.e.\ close to the critical point at $\rho'=0$.)
Note that in Ref.~\cite{Cardy98} these exponents were defined in
terms of the control parameter $\tau\sim\rho'^2$ and thus differ by
factors of two from ours. Our convention
is more appropriate for comparison with the FA model, where the
steady-state particle density is the natural control parameter.
We also note that the free propagator and hence the bare diffusion
constant $D_0$  (the constant multiplying the
term $\phi^\dag \nabla^2 \phi$ in the action) depend explicitly on
$\rho'$. In the usual RG analysis $D_0$ is set to unity, so we define the 
exponent $z$ via the scaling of typical relaxation timescales $\tau$ measured in units
of $D_0^{-1}$,
\begin{equation}
  D_0\tau \sim \xi^z \sim \rho'^{-z\nu}. 
  \label{equ:xi_exponent} 
\end{equation}
The scaling of the times $\tau$ in absolute units is then governed by
an exponent different from $z$: $D_0\propto\rho'$ gives
\begin{equation}
  \tau \sim \xi^z / D_0 \sim \rho'^{-1-z\nu}.
  \label{equ:tau_exponent}
\end{equation}

We next show that below $d=2$ the exponents are exactly 
\begin{equation}
(z,\nu,\beta)_{d<2}=(2,1/d,1).
\label{exponents_below_2}
\end{equation}
This is consistent with the exact scaling in one
dimension and also with the naive scaling estimate $\tau\sim
c^{-1-2/d}$ for the FA model~\cite{SollichReview}.
In Ref.~\cite{Cardy98}, $\nu$ and $\beta$
were given only to first order in a loop expansion: in our notation
the results were 
\begin{equation}
z=2,\qquad \beta=d\nu, \qquad \nu=2/y_\tau=1/d+\mathcal{O}(2-d)^2,
\end{equation}
for $d<2$. Here $y_\tau=2d+\mathcal{O}(2-d)^2$
is the scaling dimension of the coupling $\tau\sim \rho'^2$. 
(The relation
$\beta=d\nu$ comes from the scaling relation $\beta=z\nu/\alpha$
where $\alpha$ is the exponent for the decay of the density after
a quench to criticality, which equals $d/2$~\cite{Cardy98}.)
However, we can determine $y_\tau$ exactly since detailed balance
fixes the steady-state density to the value $\epsilon'\propto \rho'$,
so that $\beta=1$. Thus we have the exact result $\nu=1/d$, or
\begin{equation}
y_\tau = 2d,
\label{equ:ytau}
\end{equation}
and the exponents (\ref{exponents_below_2}) are exact. 
Equation~(\ref{equ:ytau}) can be confirmed by summing the
geometric series of loop corrections that contribute to 
the scaling dimension $y_\tau$.

Proceeding to the bosonic FA model, the (unshifted)
dynamical action is~\cite{Whitelam04}
\begin{eqnarray}
\fl S_\FA = 
\int\rmd^dx\,\rmd t\, \phi_{xt}^\dag \partial_t \phi_{xt}
+ \mu_0 (\phi_{xt}^\dag-1)(\phi_{xt}-\rho)(1+l_0^2\nabla^2/2)
\phi^\dag_{xt}\phi_{xt},
\end{eqnarray}
where $\rho\propto\epsilon$ is the steady state density and $\mu_0$ is
the bare coupling (with dimension [time]$^{-1}$[length]$^d$);
$l_0$ is the microscopic lengthscale as before.
The property of detailed balance manifests itself as an invariance of
the action,
\begin{equation}
S_\FA[\phi_{xt},\phi^\dag_{xt}]
 =
S_\FA[\rho\phi^\dag_{x,-t},\rho^{-1}\phi_{x,-t}].
\label{equ:S_detbal}
\end{equation}
This of course has an analogue in our earlier operator notation, where
we recognised detailed balance as the fact that $\LbFA \exp(-\beta
\hat{E})$ is Hermitian. Since $\exp(-\beta\hat{E})= \epsilon^{\sum_i
a_i^\dag a_i}$, this implies $\exp(-\LbFA t)=\epsilon^{-\sum_i a_i^\dag
a_i}\exp(-\LbFA^\dag t)\epsilon^{\sum_i a_i^\dag a_i}$ which is the
promised analogue of (\ref{equ:S_detbal}) [recall (\ref{n_exp}) and
note that $f^\dag(a,a^\dag)=f(a^\dag,a)$]. 

The duality symmetry as it is normally stated for systems with a DP
fixed point (but not necessarily with detailed balance) is~\cite{HinrichsenDP}
\begin{equation}
S_\mathrm{FA}[\phi_{xt},\phi^\dag_{xt}]
 = 
S_\mathrm{FA}[\rho(1-\phi^\dag_{x,-t}),1-\rho^{-1}\phi_{x,-t}].
\label{equ:S_dual}
\end{equation}
Like detailed balance this involves
time reversal; in terms of the Liouvillian it relates $\LbFA$
to its conjugate $\LbFA^\dag$. To arrive at the duality mapping
for the FA model, we combine the preceding two symmetries of the
action to obtain
\begin{equation}
S_\FA[\phi_{xt},\phi^\dag_{xt}]
 =
S_\FA[\rho-\phi_{x,t},1-\phi^\dag_{x,t}].
\end{equation}
In terms of operators, this symmetry now relates $\LbFA$ directly to itself,
without any conjugation. A comparison with
(\ref{equ:bos_op_duality_map}) reveals that it corresponds directly to
our earlier transformation $\tilde{W}$ from (\ref{equ:def_dt}).

Both detailed balance and duality symmetries
are preserved under renormalisation if all terms in the action are
retained. It is then crucial to follow Ref.~\cite{Cardy98} in
choosing a basis for the RG equations that reflects this 
fact. The solution is to make the transformation
defined by $\tilde{V}$ in the previous section and to write as in
(\ref{equ:bos_op_map})
\begin{equation}
\psi_{xt} = \frac{1}{2}\left(\phi_{xt} - \frac{\rho}{2}\right)
, \qquad 
\psi_{xt}^\dag = 2\left(\phi_{xt}^\dag-\frac{1}{2}\right).
\label{equ:def_psi}
\end{equation}
Hence we can write the dynamical action in terms of these new fields:
\begin{eqnarray}
\fl S_\FA[\psi,\psi^\dag] = \int\rmd^dx\,\rmd t\, \psi_{xt}^\dag \partial_t \psi_{xt}
\nonumber\\
+ \mu_0 (\psi^\dag_{xt} - 1)(\psi_{xt}-\rho/4)(1+l_0^2\nabla^2/2) (\psi^\dag_{xt} + 1)\psi_{xt}+\rho/4)
\label{equ:sfa_psi}
\end{eqnarray}
This is identical in form to $S_\AA$ as given in
Equation~(\ref{equ:saa_bare}), and so the FA model renormalises
exactly as the AA model, consistent
with their correlation functions obeying the simple relation
(\ref{equ:map_correl_bos}). The significance of the basis used is that
the resulting RG equations respect both the detailed balance and
duality symmetries of the FA model. If one makes any transform other
than (\ref{equ:def_psi}), using for example the standard shift
$\overline\phi=\phi^\dag-1$ as in Ref.~\cite{Whitelam04}, then
these symmetries are obscured and one is led to the conclusion that the
FA model is controlled by the DP fixed point between two and four dimensions.

\subsection{Simulations showing Gaussian scaling in $d=3$}
\label{sec:gauss} 

We now confirm our above exact predictions of mean-field (Gaussian) critical
exponents in $d>d_c=2$ by equilibrium simulations of the FA model 
in $d=3$. We explain how the duality relation (\ref{equ:dual_prop}) 
together with an appropiately adapted continuous time Monte Carlo 
(MC) algorithm allow us to probe citical properties well beyond the 
regime accessed in previous work. A comparison of our data with analytical
scaling forms is also given. 

A convenient observable for extracting the relaxation 
time scaling is the normalised two-point susceptibility $\chi_2(t)= \langle 
\Delta E(t) \Delta E(0) \rangle / \langle \Delta E(0)^2 \rangle$ where 
$\Delta E = E - \langle E \rangle$ denotes the fluctuation of the
energy away from its equilbrium value. Since we only consider the
stationary state dynamics of the FA model in this section, we drop the
subscript `FA,eq' on the averages. Substituting 
the simple form (\ref{equ:E}) of the energy function and using
translational invariance, $\chi_2(t)$ can be recast in the form 
\begin{equation}
  \chi_2(t) 
  = \sum_i \frac{\langle n_i(t) n_j(0) \rangle - n_\eq^2 }{n_\eq(1-n_\eq)} 
  = \sum_i \bra{0} s_i^- \rme^{-\LFA t} s_j^+ \ket{0},
  \label{equ:chi2} 
\end{equation}
with $j$ an arbitrary reference site and $n_{\eq} = \langle n_i \rangle$, 
Equation~(\ref{equ:neq}). The second equality in (\ref{equ:chi2}) follows from 
the duality relation (\ref{equ:dual_prop}). It shows that the
stationary state average defining 
$\chi_2(t)$ has a dual counterpart in the dynamics near the empty state. 
This is a tremendously useful fact: instead of having to simulate an equilibrium 
system containing hundreds or even thousands of defects we simply initialise 
with a {\em single} defect at site $j$. According to (\ref{equ:chi2}), $\chi_2(t)$ 
is then given by the probability that this state evolves under 
FA dynamics into one containing again a single defect (at an arbitrary
site $i$), that is any configuration with $E = 1$.

To measure with similar efficiency a dynamically growing lengthscale
in the FA model one can consider the mean squared displacement
associated with two-point correlations,
\begin{equation} 
  r^2_\chi(t) = \frac{1}{\chi_2(t)} \sum_i ||\bx(i)-\bx(j)||^2 
  \frac{\langle n_i(t) n_j(0) \rangle - n_\eq^2}{n_\eq(1-n_\eq)}.
  \label{equ:def_r2}
\end{equation}
Here $\bx(i)$ denotes the position vector of site $i$, 
and $j$ is again an arbitrary reference site; we set the lattice
constant to unity so that $\bx(i)\in \mathbb{Z}^d$.
Note that due to normalisation this lengthscale can exceed the
equilibrium dynamical correlation 
length, which would conventionally be extracted from the maximum of
$\chi_2(t) r^2_\chi(t)$. As in  
(\ref{equ:chi2}) we apply the duality (\ref{equ:dual_prop}) to map the stationary
state 
average in (\ref{equ:def_r2}) onto $\bra{0} s_i^- \rme^{-\LFA t} s_j^+ \ket{0}$, 
and perform simulations of the dual problem rather than the
time-consuming direct equilibrium 
simulations. 

The duality mapping increases the efficiency of an MC algorithm 
to such an extent that the limiting constraint in practical
simulations is system size rather than computational speed.
Although we initialise with just a single defect, its trajectory under FA dynamics explores 
the simulation box and must not be biased by finite size effects. For 
conventional, lattice based algorithms the required system size $N =
L^d$ can then quickly exhaust the available memory of standard computers (say 1Gb). To overcome 
this problem we used a coordinate-based variant of MC. Instead of storing the 
occupation numbers $n_i$ of $N$ lattice sites $i$ we keep track of the actual coordinates 
$\bx_a$ of each defect $a$ in the `virtual' simulation box
$\{1,2,\ldots,L\}^d$. 
The memory efficiency of this approach against a lattice based code is
$\mathcal{O}(M/N)$ if there are $M$ defects.
Hence it is useful for simulations of problems with low defect concentration, for instance, direct
equilibrium simulations at low $n_{\eq}$ or the dual dynamics near the
empty state.
In analogy to lattice based codes~\cite{BKL,NB} one can set up a `reverse lookup scheme' 
where continuous time MC steps have $\mathcal{O}(1)$ computational complexity. 
This is 
accomplished by storing the list of defect coordinates $\{ \bx_a \}_{a=1}^M$ in a hash 
table. So the problem of whether any site $\bx(i)$ in the virtual simulation box is occupied 
by a defect can be decided in $\mathcal{O}(1)$ time, just as for a lattice based code. 
This coordinate-based continuous-time MC approach is memory 
efficient while yielding computational speeds comparable to a lattice
based code. This allows us to exploit fully the simplifications
afforded by (\ref{equ:chi2}) and (\ref{equ:def_r2}).

Simulation results for branching rates $c = 10^{-3}, 10^{-4}, 10^{-5}$ and $10^{-6}$ 
are shown in Figures~\ref{fig:fa_scalingA} and \ref{fig:fa_scalingB} below. 
We have used a formally infinite virtual 
simulation box so that via (\ref{equ:chi2}) and (\ref{equ:def_r2}) we
are probing the equilibrium dynamics of the FA model {\em in the thermodynamic limit}. In other 
words, our results are guaranteed to be free of finite size effects
and can be compared directly to the scaling predictions
(\ref{equ:xi_exponent}) and (\ref{equ:tau_exponent}).
For this comparison we recall that at small defect densities the
bosonic and hard core FA models have 
similar behaviour, so that $\rho\propto\epsilon\approx c$ is
proportional to $c$. Since we know 
that $\beta=1$ exactly, it is sufficient to show that $z=2$ and $z\nu=1$ to demonstrate 
Gaussian scaling. We first verify the dynamical critical exponent $z=2$ which implies 
for the growing length scale $r_\chi(t)$, 
\begin{equation}
  r_\chi(t) \sim (D_0 t)^{1/z} \sim (c t)^{1/2}. 
  \label{equ:rchi_scaling} 
\end{equation} 
For all values of $c$ considered the data in Figure~\ref{fig:fa_scalingA} 
show a linear dependence of $r_\chi^2(t)$ on $c t$, over up to eight decades in $c t$; in 
fact our data are fully consistent with $r_\chi^2(t) = 2 d D_0 t$
where $d=3$ and $D_0 = c/2$. This value of $D_0$ confirms the expected
scaling $D_0 \sim c$; we return below to a derivation of the prefactor
in $D_0=c/2$. It should be emphasised that the kind of
spatio-temporal scaling used here determines the dynamical exponent
directly from its definition; measuring the 
ratio between exponents for the correlation length and relaxation time is a valid 
procedure only if the diffusion constant is slowly varying at criticality.
\begin{figure}
  \begin{center}
  \hfill \epsfig{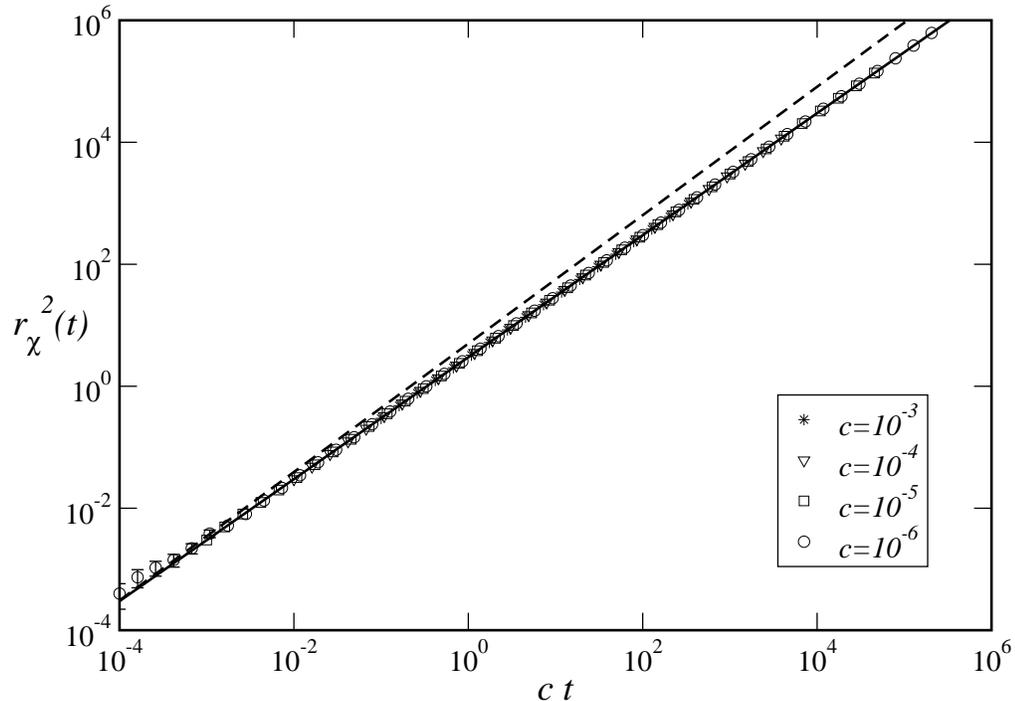} 
  \end{center}
  \caption{\label{fig:fa_scalingA} 
    Simulation data (symbols) for equilibrium dynamics in the $d=3$ FA model with branching 
    rates $c=10^{-3}, 10^{-4}, 10^{-5}$ and $10^{-6}$ obtained from a coordinate-based 
    continuous-time MC algorithm applied to the dual form 
    of $r_\chi^2(t)$, Equation~(\ref{equ:def_r2}).
    Results are averaged over $10^7,10^6,10^5$ and 
    $5 \times 10^4$ samples, respectively. 
    The dynamical lengthscale increases according to a diffusive law 
    $r_\chi^2(t) = 2 d D_0 t$ and with diffusion constant $D_0 = c/2$ (full line and symbols). 
    Error bars are significantly smaller than the symbol-size except
    where they are shown explicitly (data for $c=10^{-6}$ with $c t < 10^{-3}$). 
    The dashed line represents DP scaling of the dynamical lengthscale
    as discussed in the main text; it is inconsistent with the data.}
\end{figure}

Consider next the simulation data for $\chi_2(t)$ shown in the inset of 
Figure~\ref{fig:fa_scalingB}. Scaling arguments predict that 
$\chi_2(t)$ should be a universal function of $t/\tau$, with $\tau$ the 
relaxation time. From (\ref{equ:tau_exponent}) we expect for Gaussian exponents
\begin{equation}
  \tau(c) \sim c^{-1-z\nu} = c^{-2}.
  \label{equ:tau_c}
\end{equation}
Collapse of data for $\chi_2(t)$ under this prescription is shown in 
Figure~\ref{fig:fa_scalingB}. While there are still noticeable pre-asymptotic 
contributions at $c = 10^{-3}$ the data for $c=10^{-4}$ and in particular 
$c=10^{-5}, 10^{-6}$ seem to have converged to the final scaling form of $\chi_2(t)$ to 
within our numerical accuracy. To confirm the critical scaling in more detail we next give
a theoretical analysis that predicts the precise shape of $\chi_2(t)$
in the critical limit $c \to 0$.
\begin{figure}
  \begin{center}
  \hfill \epsfig{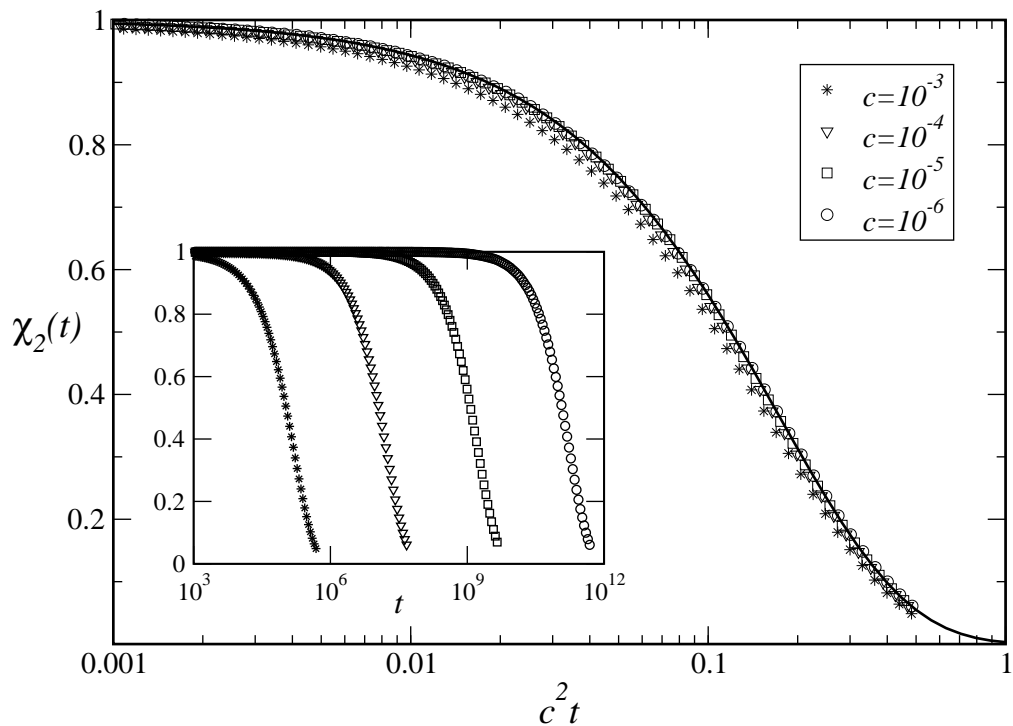} 
  \end{center}
  \caption{\label{fig:fa_scalingB} 
    Simulation data obtained from the dual representation (\ref{equ:chi2}) of 
    the two-point susceptibility $\chi_2(t)$ 
    and with the same parameters as in Figure~\ref{fig:fa_scalingA}. 
    The inset shows $\chi_2(t)$ versus $t$. 
    At each $c$ data are collected up to times $t = 0.5/c^2$, that is up to $t=0.5 \times 
    10^{12}$ for the lowest $c$. The main plot demonstrates data collapse with
    the Gaussian scaling of the relaxation time $\tau \sim c^{-2}$. The full line 
    represents our analytical prediction for the $c \to 0$ scaling of $\chi_2(t)$, 
    Equation~(\ref{equ:chi2scaling}); there are no fit parameters. Error 
    bars are largest for the data at $c=10^{-6}$; even there we
    estimate a relative error below $1\%$.}
\end{figure}

\subsection{Scaling analysis for two-point functions in $d>2$}
\label{sec:scaling}
We now show that the $c \to 0$ limit of the equilibrium two-point susceptibility 
for FA models in $d>2$ dimensions is 
\begin{equation}
  \chi_2(t) = \exp(-\kappa_d c^2 t),
  \label{equ:chi2scaling} 
\end{equation}
where $\kappa_d = d \left[ \mathcal{P}_d(2,0,\ldots,0) + 2 (d-1) 
\mathcal{P}_d(1,1,0,\ldots,0) \right]$, and $\mathcal{P}_d(\bm{x})$ is the 
probability that a pair of random walkers with initial separation
$\bm{x}$ will never meet: this function is considered in
\ref{app:p_d}. Equation~(\ref{equ:chi2scaling}) is consistent in particular with the
scaling expectation that $\chi_2(t)$ should be a 
function of $t/\tau$ with $\tau \sim c^{-2}$. In $d=3$ and substituting the survival 
probabilities (\ref{equ:p200}) and (\ref{equ:p110})
one has $\kappa_3 \approx 5.80961$. The exact $c \to 0$ scaling form (\ref{equ:chi2scaling}) 
for this dimensionality is shown in 
Figure~\ref{fig:fa_scalingB} and is in perfect agreement  
with our simulation data.
The remainder of this subsection is devoted to the derivation of
(\ref{equ:chi2scaling}); we return to more general discussion in
Section~\ref{sec:pers}.

A systematic analysis of the dynamics in the critical limit $c\to 0$ requires
a careful distinction between the timescales involved. We will need
unscaled time $t$ as well as the scaled time variables $x=ct$ and
$y=c^2t$ whose $\mathcal{O}(1)$ increments correspond to time
intervals $\delta t$ of $\mathcal{O}(c^{-1})$ and $\mathcal{O}(c^{-2})$, respectively.
For time intervals $\delta t=\mathcal{O}(1)$ the limit $c \to 0$ is
clear: the rate $c$ for branching processes $1_j 0_k \to 1_j 1_k$, where $k$ 
is a nearest neighbour (NN) of $j$, then vanishes. Only the coagulation processes
$1_j 1_k \to 1_j 0_k$ or $1_j 1_k \to 0_j 1_k$ can then take place, each occuring with 
rate unity, and the lifetime of excited states like $1_j 1_k$ that
contain two (or more) NN defects is $\mathcal{O}(1)$. 

The dynamics over intervals with $\delta x = \mathcal{O}(1)$, i.e. $\delta t =
\delta x/c = \mathcal{O}(c^{-1})$, is rather different. Processes
involving branching events are then possible. However, after such an
initial event the rate for reverting to the original state is
$\mathcal{O}(1)$, while the rate for an additional branching event
is $\mathcal{O}(c)$. This implies that the probability for the latter
to occur first is $\mathcal{O}(c)$. Since the rate for the first branching event
is already $\mathcal{O}(c)$, the effective rate for a sequence of two
branching events is then $\mathcal{O}(c^2)$. On our
$\mathcal{O}(c^{-1})$ timescales this type of process can be neglected, and
we need only concern ourselves with processes involving a single branching event.
One possible process is then $1_j 0_k \to 1_j 1_k \to 0_j 1_k$, with $j,k$ NNs. 
Its rate is $c \times \frac{1}{2}$, $c$ being the branching rate
and $\frac{1}{2}$ the probability for the particular, subsequent
coagulation (rather than $1_j 1_k \to 1_j 0_k$, which would take us back to
the initial state). The probability for this process to occur during the
interval $\delta t$ is therefore $\frac{1}{2} c \delta t = \frac{1}{2} \delta x$. 
Note that as the lifetime of excited states is $\mathcal{O}(1)$ as argued above, 
we have vanishing probability $\mathcal{O}(c)$ of finding the system in an 
excited state $1_j 1_k$ at any given moment in time. Therefore, 
in the limit $c \to 0$, the intermediate excited step of our process
becomes invisible and we have {\em effective diffusion} $1_j 0_k \to 0_j 1_k$ 
with probability $\frac{1}{2} \delta x$.

The second type of process allowed on the $\mathcal{O}(c^{-1})$
timescale is the excitation of a defect at a site that has more than
one NN defect, followed by a cascade of
coagulation events leading to another configuration where all defects
are isolated.
Consider first the case where the new defect, at site
$k$ (say), has two NN defects at sites $j$ and $l$, so that our
process leads from $1_j 1_k 1_l$ to $1_j 0_k 0_l$ or $0_j 1_k 0_l$ or
$0_j 0_k 1_l$. The corresponding probabilities are $\frac{1}{4} \delta
x, \frac{1}{2} \delta x, \frac{1}{4} \delta x$, respectively: after
the initial excitation (probability $2c\delta t=2\delta x$) at site
$k$, there is a probability of 1/2 that the defect at site $k$, which
has twice the down-flip rate of those at $j$ and $l$, will not flip
down first. (If it does, we have returned to the original
configuration and can ignore the process.) There is then probability
1/2 that $j$ will flip before $l$, and probability 1/2 for each of the
remaining defects to flip first, resulting in the overall probabilities
given above.

To summarise the dynamics on the $O(c^{-1})$ timescale, we have
diffusion $1_j 0_k \to 0_j 1_k$ with rate $\frac{1}{2}$ per interval of
rescaled time $x$. The processes $1_j 0_k 1_l \to 1_j 1_k 1_l \to
\ldots$ discussed 
above can then be represented consistently as produced by a diffusion
event followed by (in the limit $c\to 0$) instantaneous
coagulation. For instance, $1_j 0_k 1_l \to 1_j 0_k 0_l$ amounts to
diffusion $0_k 1_l \to 1_k 0_l$ (with rate $\frac{1}{2}$) followed by
coagulation $1_j 1_k \to 1_j 0_k$ (which has probability $\frac{1}{2}$ of
occurring before $1_j 1_k \to 0_j 1_k$), giving the overall rate
$\frac{1}{2} \times \frac{1}{2} = \frac{1}{4}$ obtained above. One can
check that the same decomposition into diffusion and instantaneous
coagulation also holds for processes involving an initial excitation
at a site with more than two NN defects.

Our discussion so far leads to the following conclusions. Clearly 
any single-defect state $s_j^+ \ket{0}$ is blocked on the $\mathcal{O}(1)$ 
timescale. Therefore $\chi_2(t)=1$ and $r_\chi^2(t)=0$ in equilibrium and for 
$\mathcal{O}(1)$ times, according to the dual representations (\ref{equ:chi2}) and 
(\ref{equ:def_r2}). But on the $\mathcal{O}(c^{-1})$ timescale the
defect diffuses away from its starting site $j$. This still predicts
$\chi_2(t) = 1$ since diffusion conserves the number of defects $E = 1$.
The quantity $r_\chi^2(t)$ in dual representation, on the other hand, now 
measures the mean squared displacement of a single, diffusing defect which is 
given by $r_\chi^2(t) = 2 d D_0 t$. Our result that the diffusion rate 
is $\frac{1}{2}$ in time units of $x=c t$ tells us that $D_0=c/2$. This is the 
exact low-density scaling of $D_0$ and precisely what we found in our
simulations; compare Figure~\ref{fig:fa_scalingA}. 

Let us now turn to dynamics on the 
$\mathcal{O}(c^{-2})$ timescale, corresponding to $\delta y=c^2 \delta
t = \mathcal{O}(1)$. During such a time interval there 
is now a nonzero probability for the occurence of processes involving two successive
branching events, where one defect is first excited next to an existing one
and another defect is then created on a NN site to either of the other two.
In order to determine 
the fate of this defect triple $1_j 1_k 1_l$ we consider its evolution
on the faster 
$\mathcal{O}(c^{-1})$ timescale. This argument is analogous to the one
above, where a defect pair is created on the $\mathcal{O}(c^{-1})$
timescale but we have to look at $\mathcal{O}(1)$ times 
to determine its relaxation. In further analogy we note that an arbitrarily 
small but nonzero increment $\delta y$ on the $\mathcal{O}(c^{-2})$ timescale corresponds 
to an infinite increment $\delta x = \delta y/c$ of $\mathcal{O}(c^{-1})$ time 
as $c \to 0$. The following possibilities then arise: because NN defects 
coagulate instantaneously on the $\mathcal{O}(c^{-1})$ timescale there is a 
probability of $\frac{1}{2}$ for immediate relaxation $1_j 1_k 1_l \to 1_m$ 
where $m=j$ or $k$ or $l$. Subsequently the defect $1_m$ diffuses with 
diffusion rate $\frac{1}{2}$ for a time $\delta x \to \infty$, moving
arbitrarily far from its initial position. With the remaining
probability of $\frac{1}{2}$ an instantaneous relaxation of the middle
defect takes place, $1_j 1_k 1_l \to 1_j 0_k 1_l$.
The defects $1_j$ and $1_l$ can now diffuse independently for an
effectively infinite interval $\delta x$ of
$\mathcal{O}(c^{-1})$-time; if they do not coagulate in the process,
their distance grows without bound and we can say the original defect
has branched irreversibly. We derive in \ref{app:p_d} the
`survival probability' $\mathcal{P}_d(\bx)$ for this outcome; here $\bx = \bx(l) - \bx(j)$
is the initial separation of the diffusing defects.

We can now assemble the probability 
$\lambda \delta t$ that during a time interval $\delta t = \delta y/c^2$ 
a single defect irreversibly branches into two defects. 
Starting from a single defect there is a rate $c$ for branching on a 
given NN site. Since (on a hypercubic lattice in $d$ dimensions) there
are $2d$ such states, the overall branching rate is $2 d c$. The
probability for a second branching event on a neighbouring site to take place before either
of the two possible relaxations
back to a single defect is $c/2$. A cluster of two NN defects 
has $4d-2$ NN sites, two of which lead to a linear and $4(d-1)$ to an angled 
defect triple $1_j 1_k 1_l$. Altogether the rate for creation of a linear triple is 
$2 d c^2$ while it is $4 d (d-1) c^2$ for an angled one. In either
case, we need to multiply by the probability $\frac{1}{2}$ of the
middle defect $k$ relaxing first, leading to a pair of next nearest 
neighbour (NNN) defects. In terms of the eventual survival probabilities 
$\mathcal{P}_d(\bx)$ of this pair we thus obtain 
\begin{equation} 
  \lambda = d \left[ \mathcal{P}_d(2,0,\ldots,0) + 
   2 (d-1) \mathcal{P}_d(1,1,0,\ldots,0) \right] c^2.
  \label{equ:lambda}  
\end{equation} 
This is a nontrivial result. To clarify its intuitive meaning,
note that we are considering initially a single defect $s_j^+
\ket{0}$. The following trajectories are then possible during a time
interval $\delta t = \delta x/c = \delta y/c^2$ on the $\mathcal{O}
(c^{-2})$ scale: (i) no branching occurs. The defect diffuses with rate $\frac{1}{2}$ 
for a time $\delta x = \delta y/c \to \infty$ in the $c \to 0$ limit. The defect 
at the beginning and end of the interval $\delta t$ are then completely decorrelated. (ii) With 
probability $d(2d-1)c^2 \delta t = d(2d-1) \delta y$ the defect branches 
into a pair of NNN defects. Again, on the $\mathcal{O}(c^{-1})$
timescale this pair has an infinite time $\delta x$ available to
diffuse through the system. (ii.a) There is a finite probability 
that the pair coagulates during this diffusive motion. In this case
we have within the time interval $\delta t$ a `bubble' in the
space-time diagram of the defect trajectories~\cite{GCTheory}, where the initial
defect separates into two but these re-coagulate shortly
afterwards. The temporal extent of this bubble, i.e.\ the time during
which the two defects exist, is
$\mathcal{O}(c^{-1})$. The probability of detecting such a bubble on the 
$\mathcal{O}(c^{-2})$ time scale is therefore vanishingly small in the 
limit $c \to 0$. This is in analogy to excited states becoming invisible on the 
$\mathcal{O}(c^{-1})$ timescale. Hence the trajectories (i) and (ii.a) cannot be 
distinguished on the $O(c^{-2})$ timescale. (ii.b) The defects may diffuse forever 
($\delta x \to \infty$) without encountering each other; this means that the trajectory branches 
irreversibly on the $\mathcal{O}(c^{-2})$ time scale. Due to the existence of bubbles the 
probability for this event, $\lambda \delta t$, is renormalised
relative to the `bare' probability $d(2d-1)c^2 \delta t$ for an
initial branching event. Each defect in the resulting pair 
has travelled an infinite distance during $\delta t$ and so completely
decorrelates from the initial defect.

We can now make predictions for the dynamics on 
the $\mathcal{O}(c^{-2})$ timescale. First we note that in $d=1,2$ the survival 
probabilities $\mathcal{P}_d(\bx)$ vanish, see \ref{app:p_d}. Hence 
$\lambda = 0$ and defect trajectories do not branch on the $\mathcal{O}(c^{-2})$ 
timescale. However, in $d>2 $ the $\mathcal{P}_d(\bx)$ are finite,
and hence so is $\kappa_d=\lambda/c^2$. Now $\chi_2(t)$ is just the probability
that the number of defects has not increased during the time interval
$t$, which means that no irreversible branching processes have taken
place. The rate for occurrence of the latter being $\lambda$, it follows that
$\chi_2(t)=\exp(-\lambda t)=\exp(-\kappa_d c^2 t)$. This completes our
derivation of Equation~(\ref{equ:chi2scaling}).

One can go further and extract the probability $p_t(E)$ of having $E
\geq 1$ at time $t$. Since each irreversible branching event produces
infinitely separated defects, these will then continue to branch {\em
independently} in the same way. Thus, if $E$ defects are present, the rate
for generating an additional one by irreversible branching is
$E\lambda$. This gives the master equation
\begin{equation}
  \partial_t \, p_t(E) = (E-1) \, \lambda \, p_t(E-1) - E \, \lambda \, p_t(E).
  \label{equ:p_E_system} 
\end{equation}
This can be solved straightforwardly, for example by
$\mathcal{Z}$-transform, 
to give
\begin{equation}
  p_t(E) = \rme^{-\lambda t} \left( 1 - \rme^{-\lambda t} \right)^{E-1}.
  \label{equ:p_E} 
\end{equation}  
The number of defects thus has an exponential distribution; the most likely outcome is $E=1$
at any time and has probability $p_t(1) = \chi_2(t)=\rme^{-\lambda
t}$. The average number of defects, on the other hand, grows
exponentially as $\langle E \rangle_t = 
\rme^{+\lambda t}$. We have checked in our simulations the full form
of $p_t(E)$ (data not shown), and found excellent agreement. The
exponential increase in the number of defects limits the time range
that can be conveniently simulated; the lowest value of $\chi_2(t)$
that can be measured reliably is of the order of the inverse of the
maximum number of defects one is prepared to track.

\subsection{Persistence functions and discussion of earlier data}
\label{sec:pers}

In order to characterise a dynamical fixed point in the presence of a 
time-reversal symmetry, we must determine three independent exponents. 
One such set is $(z,z\nu,\beta)$. Our simulations as well as the scaling 
analysis have confirmed the RG prediction that $z=2$ and $z\nu=1$; further, $\beta=1$ 
is known rigorously from detailed balance. Consequently the scaling of the FA 
model in $d=3$ is Gaussian. 

In Ref.~\cite{Whitelam04}, the authors found that the directed percolation (DP) fixed 
point is relevant to the FA model in three dimensions and at low temperatures. 
This conclusion seems unsatisfactory: we find no evidence for fluctuation 
corrections in our simulation data. Further, and more importantly, the DP fixed 
point is characterised by a diverging static lengthscale $\xi$: for a process 
in the DP universality class, one expects
$\langle n_i(0) n_j(0) \rangle_\mathrm{DP} - \langle n \rangle^2_\mathrm{DP}$
to be a scaling function of $||\bx(i)-\bx(j)||/\xi$ where $\xi$ diverges
at criticality. On the other hand, detailed balance with respect to
the non-interacting energy function (\ref{equ:E}) tells us that
$\langle n_i(0) n_j(0) \rangle_\FA - \langle n \rangle_\FA^2
\propto \delta_{ij}$ in the FA model. The absence of this diverging lengthscale 
in the FA model indicates that the underlying physics is different from that
of the DP fixed point. Finally, we note that the exact result $\beta=1$
arises naturally from our RG treatment: this situation would appear more satisactory than the 
argument of Ref.~\cite{Whitelam04} that the exponents $z$ and $\nu$ should be given by 
their DP values while $\beta$ is independently
fixed to the non-DP value $\beta=1$ by detailed balance.

\begin{figure}
  \hfill \epsfig{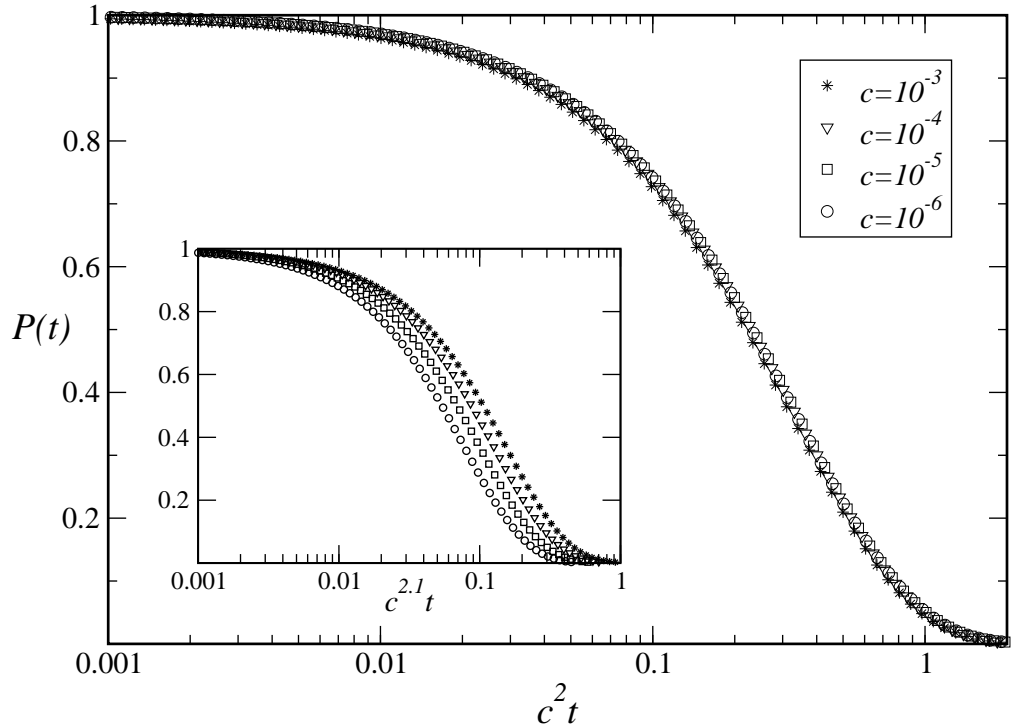}
  \caption{\label{fig:fa_persistence} 
    Simulation data (symbols) for the equilibrium persistence function $P(t)$ in the $d=3$ 
    FA model with branching rates $c=10^{-3}, 10^{-4}, 10^{-5}$ and $10^{-6}$. The main 
    panel shows data collapse for Gaussian scaling of the relaxation time $\tau \sim 
    c^{-2}$. The inset demonstrates that the data are inconsistent 
    with DP scaling $\tau_\mathrm{DP} \sim c^{-2.105}$. 
    We used the coordinate-based continuous-time MC algorithm to
    measure $P(t)$ in direct
    equilibrium simulations; since $P(t)$ is effectively a correlation
    function of events at a continuous range of times, the duality
    relations for two-time correlations cannot be used. The size of the
    virtual simulation box is $64^3, 200^3, 640^3$ 
    and $2000^3$, respectively, which should be sufficient to avoid 
    finite-size effects. 
    Results are averaged over $1000, 100, 10$ and $3$ repeats, again 
    in order of decreasing $c$. We expect relative errors of no more than $1\%$ in 
    the data shown.}
\end{figure}

We argue that the conclusions of Ref.~\cite{Whitelam04}
regarding the DP fixed point are artefacts of an RG treatment that does
not respect the presence of detailed balance and of the duality
symmetry of the FA model; see also the comments after Equation~(\ref{equ:def_psi}).
To remedy this, we have shown explicitly that writing the action as in~(\ref{equ:sfa_psi})
allows one to perform the RG analysis in a way that preserves these
symmetries. We attribute the apparent fluctuation effects in 
the data for the relaxation time of Ref.~\cite{Whitelam04}, which
were derived from the persistence function $P(t)$, to a combination of pre-asymptotic 
corrections in $c$ and possible finite size effects. It should be
emphasised that pre-asymptotic corrections are substantial: our
simulations for $\chi_2(t)$ show that branching rates of $c=10^{-3}$ are still too
large to see the true critical scaling. Similar comments apply to the
persistence function: we do find Gaussian scaling as shown in
Figure~\ref{fig:fa_persistence}, but to see this clearly requires very
small $c$. Note that in order to obtain data for $c = 10^{-5}$ and $c = 10^{-6}$ 
we used virtual simulation boxes of size $640^3$ and $2000^3$, 
respectively~\footnote{
    While our coordinate-based MC algorithm is extremely memory 
    efficient (in a $2000^3$ lattice and at $c = 10^{-6}$ there are on average only 
    $8000$ defects) keeping track of the persistence status is a problem for large 
    lattices. We associate single bits with the persistence status of
    each lattice site
    so that we can track persistence in lattices up to $1000^3$ using a memory block 
    of 125Mb. At $c=10^{-6}$ we split the $2000^3$ virtual simulation box into eight 
    $1000^3$ blocks but only track persistence in four of them to
    limit the memory requirement to 500Mb.
    }. 
These cannot be reached by conventional lattice-based codes
so that the data of Ref.~\cite{Whitelam04} were of necessity
taken from smaller systems, with potentially significant finite size
effects.
Although the DP exponents $z_\mathrm{DP}=1.9$ and $\nu_\mathrm{DP}=0.58$ 
in $d=3$~\cite{HinrichsenDP} differ only slightly from the Gaussian ones,
our data clearly allow us to rule them out: the dashed 
line in Figure~\ref{fig:fa_scalingA} represents 
DP scaling of the dynamical correlation length and is inconsistent 
with our data. The inset in Figure~\ref{fig:fa_persistence} demonstrates
similarly that the persistence functions $P(t)$ do not collapse 
when plotted against $t/\tau_\mathrm{DP}$ with the DP scaling $\tau_\mathrm{DP} 
\sim c^{-2.105}$; a rather similar picture -- thus not shown --  is 
obtained when plotting the two-point susceptibility $\chi_2(t)$, 
Figure~\ref{fig:fa_scalingB}, against $t/\tau_\mathrm{DP}$. 

We conclude that the DP fixed point is irrelevant for the FA model,
for the same reason that it is irrelevant for the AA model and for parity 
conserving models of branching and annihilating random walks: these 
models possess extra symmetries that must be preserved in RG calculations
and, in the case of the FA model, lead to Gaussian scaling.
This conclusion parallels that of Cardy and T\"auber~\cite{Cardy98},
who showed that an early paper~\cite{Grassberger89}
on the parity conserving
reaction-diffusion system ($A\to3A, 2A\to0$) had produced a similar
erroneous conclusion that the upper critical
dimension was four and the exponents those of DP.

\section{Conclusion}
\label{sec:conc}

To summarise, we showed in Section~\ref{sec:map} that the FA and AA
models with hard core exclusion share the same
correlation functions (at equilibrium, and considering a single
time-difference). This was established by means of an exact
mapping at the level of the master equation. An important
generalisation which includes additional diffusive processes
demonstrated that the same mapping connects more generally the
reaction-diffusion models with reversible coagulation
$A+A\leftrightarrow A$ and reversible annihilation $A+A\leftrightarrow
0$. Further, we showed in Section~\ref{sec:bos} that the bosonic
versions of the FA and AA models are appropriate effective theories
for the low temperature limits of the hard core models and have
analogous symmetries and relations between each other.
Finally, in
Section~\ref{sec:crit} we discussed the critical properties of the
bosonic models using renormalisation group arguments. Implementing the
mapping at the level of the field-theoretic action showed that the FA
(and more generally $A+A\leftrightarrow A$) model renormalises like
the AA (or $A+A\leftrightarrow 0$) model. We find that the directed percolation fixed
point is irrelevant to the FA model, because of the presence of
detailed balance and of an additional hidden symmetry inherited from
the parity symmetry of the AA model. Instead, the $A+A\leftrightarrow
A$ model and its special case, the FA model, have upper critical
dimension two; detailed balance together with the two exactly known
scaling exponents is sufficient to find all exponents exactly also in
$d<2$.

From the point of view of reaction-diffusion systems and, more
generally, non-equilibrium stochastic models, the most significant
outcome of this work is the result that a hidden symmetry suppresses
fluctuations in the $A+A\leftrightarrow A$ model and lowers its upper
critical dimension to $d_c=2$. The mapping to $A+A\leftrightarrow 0$
can, however, also be used to more quantitative purposes. For example,
it enables one to calculate new exact results for two-time
non-equilibrium correlation and response functions in $d=1$; we will
report on these shortly~\cite{MaySol06}. The results of such an analysis are
instructive also more generally with regard to
non-equilibrium fluctuation-dissipation relations for activated
dynamics~\cite{FA-FDT}. In fact the FA model is an almost paradigmatic example of
such dynamics, given that any evolution away from a
metastable state containing only isolated defects requires the thermal
excitation of additional defects. 

From a different angle, one may ask what our results have to say about
the usefulness of the FA model for capturing the qualitative behaviour
of structural
glasses~\cite{SollichReview,GCTheory,Whitelam04}. We have seen
that in the physically relevant case of three spatial dimensions,
fluctuation effects at low defect densities are of a classical
(mean-field, Gaussian) nature. Nevertheless, the models will still
exhibit a degree of dynamical heterogeneity. Violations of the Stokes-Einstein
relation~\cite{Jung-SEB} may also persist, but will be at most by a
constant (rather than diverging) factor as $c\to 0$; this is consistent with simulation
results~\cite{JPG_private}. In summary, `glassy' effects will be
present, but probably rather weak. This is consistent with the fact
that FA models also have relatively benign, Arrhenius-type increases of
relaxation time scales at low temperature: $\tau\sim c^{-2}\sim
\exp(2/T)$ in $d>2$ as we saw above. These models are therefore suitable at best
for modelling for what are known as strong glasses. For fragile glasses
with their super-Arrhenius timescale divergences,
models with facilitation by more than one spin -- or with directed
constraints -- will inevitably have to
be used. Their much more cooperative
dynamics~\cite{SollichReview} continues to make them physically
attractive models for understanding non-trivial aspects of glassy
dynamics.

\ack
We thank L.\ Berthier, P.\ Calabrese, J.\ Cardy, J.-P.\ Garrahan, D.\ Reichman, R.\ Stinchcombe,
U.\ T\"{a}uber and S.\ Whitelam for inspiring discussions. RLJ was supported by
EPSRC grant no. GR/R83712/01.

\appendix 

\section{Large-$S$ expansion}
\label{sec:large_S}

To see mathematically the equivalence between the bosonic and hard core
models in the limit of small particle densities, one can replace the spin-1/2
operators of the hard core case by their spin-$S$ analogues and
perform a formal large-$S$ expansion, valid for states with small density.
For example, we can define a new Master operator by generalising
$\LFA$ from (\ref{LFA_def}) to $S>1/2$,
as follows:
\begin{equation}
\mathcal{L}_\mathrm{SFA} = \sum_{\langle ij\rangle}
\left[\frac{S+S^z_j}{2S}(S^+_i-\sqrt{2S})\frac{S-S^z_i}{2S}
(S^-_i-\sqrt{2S}c) + (i\leftrightarrow j)\right],
\end{equation}
where $S^z_i$ etc are operators in the spin-$S$ algebra. In the spin-half
case we have $S=1/2$, $(S+S^z_j)/(2S)=s^+_j s^-_j$, $(S-S^z_i)/(2S)=s^-_is^+_i$ and so
recover immediately $\mathcal{L}_\mathrm{SFA}=\LFA$.

The Master operator $\mathcal{L}_\mathrm{SFA}$
describes a system in which the number of particles on
each site is restricted to the range $0\leq n_i\leq 2S$; the particle
number operators are $\hat{n}_i=S_i^z+S$.
Configurations $\{n_i\}$ are mapped onto kets $\prod_i (S^+_i/\sqrt{2S})^{n_i} |0\rangle$ where
$|0\rangle$ is the empty state as before; since $\hat{n}_i|0\rangle=0$
is equivalent to $S_i^z|0\rangle=-S|0\rangle$ this state is, in spin
language, fully polarised in the $(-z)$ direction. Probabilities for transitions 
between states in some time interval $t$ are then given by
\begin{equation}
P_{\{n_i'\}\leftarrow\{n_i\}}(t)=\langle 0 | 
\left[ \prod_i
\frac{1}{\Gamma_{S,n_i'}} \left(\frac{S_i^-}{\sqrt{2S}}\right)^{n_i'}
\right] \rme^{-\mathcal{L}_\mathrm{SFA}t} \left[ \prod_i
\left(\frac{S^+_i}{\sqrt{2S}}\right)^{n_i} \right]
| 0 \rangle
\label{transn_prob_general_S}
\end{equation}
where we have introduced the coefficients 
$\Gamma_{S,n}=(2S)^{-n}\langle0|(S_i^-)^{n}(S^+_i)^{n}|0\rangle$
for ease of writing. These obey the recursion
$\Gamma_{S,n+1}=\Gamma_{S,n}(n+1)[1-n/(2S)]$, yielding explicitly
$\Gamma_{S,n}=n!(2S)!/[(2S-n)!(2S)^n]$. 
From (\ref{transn_prob_general_S}), conservation of probability requires that
\begin{equation}
\langle 0 | \left[ \prod_i \sum_{n_i=0}^{2S}
\frac{1}{\Gamma_{S,n_i'}} \left(\frac{S_i^-}{\sqrt{2S}}\right)^{n_i'}
\right] \mathcal{L}_\mathrm{SFA} =0
\end{equation}
which can be verified by direct calculation. We identify this left eigenstate
as the projection state; it is analogous to $\langle e |$ and $\langle \tilde e|$ for
hard core and bosonic models, respectively.

Using $S^+_i(S^+_i/\sqrt{2S})^n|0\rangle =
\sqrt{2S}(S^+_i/\sqrt{2S})^{n+1}|0\rangle$ and $S^-_i(S^+_i/\sqrt{2S})^n|0\rangle =
\sqrt{2S} \times n[1-n/(2S)](S^+_i/\sqrt{2S})^{n-1}|0\rangle$ one easily
checks that the microscopic rates in the model defined by
$\mathcal{L}_\mathrm{SFA}$ are
\begin{equation} 
  \begin{array}{cccl}
    n_i n_j & \rightarrow & (n_i+1) n_j, & \quad \mbox{rate } 
    c n_j [1-n_i/(2S)]
            , \\
    (n_i+1) n_j & \rightarrow & n_i n_j, & \quad \mbox{rate } 
        n_j (n_i+1)[1-n_i/(2S)]^2
  \end{array} 
\label{S_rates}
\end{equation}
These obey detailed balance with respect to the stationary state
\begin{equation}
P(\{ n_i \}) \propto \prod_i \frac{c^{n_i}}{\Gamma_{S,n_i}}\ .
\label{S_stationary}
\end{equation}

We have defined a family of interpolating models with increasing $S$ 
that allow us to gradually remove the hard core constraints. 
We will now use a large-$S$ expansion
to show that the models without constraints coincide with the bosonic models
defined above. As long as there is no qualitative change in behaviour on
increasing $S$ we therefore expect the bosonic models to be suitable
effective theories for the low temperature (small $c$)
behaviour of the hard core ones. An example of a qualitative change
that would render the large-$S$ expansion 
invalid is a transition to a quantum disordered state as $S$ is
reduced~\cite{SachdevQPT}: there is clearly no such singular behaviour
here. Indeed, in our case the stationary states (\ref{S_stationary}) of 
the models are known for general $S$. Bearing in mind that
$\Gamma_{S,0}=\Gamma_{S,1}=1$, they are all of effectively the same form
if $c$ is small so that only states with $n_i=0,1$ have significant
probability.

Our claim that the above interpolating model becomes equivalent to the
bosonic one in the limit $S\to\infty$ can be confirmed directly from
(\ref{S_rates}): as long as $c$ is small so that the relevant values
of $n_i$ stay small compared to $2S$, the large-$S$ limit of the
transition rates gives the bosonic model
(\ref{bosonic_FA}). Correspondingly, the stationary state
(\ref{S_stationary}) becomes the bosonic one in this limit since
$\Gamma_{S,n}\to n!$.

More formally, one can establish the large-$S$ limit of our
interpolating model by looking at the Hermitian version of the Liouvillian.
Using detailed balance this is defined as
$H_\mathrm{SFA}=[\prod_i c^{-\hat{n}_i/2}]\mathcal{L}_\mathrm{SFA}[\prod_i
c^{\hat{n}_i/2}]$ or explicitly
\begin{equation}
H_\mathrm{SFA} = \sum_{\langle ij\rangle}
\left[\frac{S+S^z_j}{2S}(S^+_i-\sqrt{2Sc})\frac{S-S^z_i}{2S}
(S^-_i-\sqrt{2Sc}) + (i\leftrightarrow j)\right]
\end{equation}
Then we can use the Holstein-Primakov representation~\cite{HP} 
\begin{equation}
S^z_i = a^\dag_i a_i - S, \qquad S^+_i=a^\dag_i(2S-a^\dag_i a_i)^{1/2}
\end{equation}
and take the large-$S$ limit by approximating $2S-a^\dag_i
a_i=2S-\hat{n}_i\approx 2S$ everywhere. This assumes again that $c$ is
small enough so that all relevant states have particle numbers $n_i\ll
2S$ at each site. In spin language, all the states of interest
are then localised on a small part of the surface of the spin sphere, and
the non-trivial structure of the spin algebra can be neglected
in favour of a simple bosonic one via $S^+_i\approx
\sqrt{2S}a^\dag_i$.
Taking the $S\to\infty$ limit as explained, we get
\begin{equation}
H_\mathrm{SFA} \simeq \sum_{\langle ij\rangle} [ a^\dag_j a_j 
( a^\dag_i - \sqrt{c} ) (a_i - \sqrt{c}) +(i\leftrightarrow j) ].
\end{equation} 
This coincides with the Hermitian form (\ref{HbFA})
of the Liouvillian of the bosonic FA model as claimed, with the expected correspondence
$\epsilon=c$. An exactly analogous procedure can be applied to
construct a family of models that interpolates smoothly between the
hard core and bosonic AA models. 
We therefore expect that the bosonic FA and AA models will be appropriate
effective theories for their hard core counterparts at small particle densities.

\section{Random walk survival probabilities} 
\label{app:p_d}

For the scaling analysis of equilibrium correlation functions given in 
Section~\ref{sec:scaling} we required particular random walk survival probabilities; 
these are derived in the following. Consider a pair of diffusing
defects. We can think of these as random walkers on a $d$-dimensional hypercubic
lattice $\mathbb{Z}^d$; whenever they occupy NN sites,
where their positions have distance $||\bx_2 - \bx_1||=1$, they coagulate
instantaneously. We are interested in the probability $\mathcal{P}_d(\bx)$ 
that the walkers survive to infinite time, i.e.\ never
coagulate. This survival probability depends on the spatial
dimensionality $d$ and on the initial separation $\bx = \bx_2 -\bx_1$
of the walkers. The distance vector $\bx_2 -\bx_1$ also performs a
random walk, with twice the effective diffusion constant. The problem
is therefore to calculate the probability $\mathcal{P}_d(\bx)$ that a
random walker starting from position $\bx$ will never reach one of the
NN sites of the origin. If we picture these sites as absorbing, then
$\mathcal{Q}_d(\bx)=1-\mathcal{P}_d(\bx)$ is the probability that the walker
is absorbed eventually; for absorption at the origin itself, these
quantities are well known.

The key insight is that, in its first step, the walker randomly
moves to one of the NN sites of $\bx$; we write these as $\by\in
N(\bx)$. The absorption probability starting from $\bx$ is therefore
the average of those that would be obtained when starting from any of
these NN sites:
\begin{equation}
\mathcal{Q}_d(\bx) = \frac{1}{2d}\sum_{\by\in N(\bx)}
\mathcal{Q}_d(\by).
\end{equation}
The only exception to this relation is the case where $\bx$ itself is
already an absorption site so that $\mathcal{Q}_d(\bx)=1$. We can
correct for this by adding a source term at these sites; the latter
then has to be set at the end of the calculation to give the correct
values of $\mathcal{Q}_d(\bx)$ at the absorption sites. Since all
lattice directions are equivalent, all $2d$ source terms will be
equal and we can write 
\begin{equation}
\mathcal{Q}_d(\bx) = \frac{1}{2d}\left[
v\delta_{\bx,N(\boldsymbol{0})} + \sum_{\by\in N(\bx)} \mathcal{Q}_d(\by) \right].
\end{equation}
For the Fourier components $Q_{\bk} = \sum_{\bx} \mathcal{Q}_d(\bx)
\rme^{-\iu \bk \cdot \bx}$ this gives
\begin{equation}
Q_{\bk} = v\, \frac{\sum_\alpha \cos k_\alpha}{d-\sum_\alpha \cos
k_\alpha}
= v \left[-1+ d \int_0^\infty \rmd\tau\, \rme^{-\tau(d-\sum_\alpha \cos k_\alpha)}\right],
\end{equation}
where $\alpha=1,\ldots,d$ labels the lattice directions. The second,
integral form of the result makes the reverse Fourier transform
simple:
\begin{equation}
\mathcal{Q}_d(\bx) = 
v \left[-\delta_{\bx,\boldsymbol{0}} + d \int_0^\infty \rmd\tau\,
\rme^{-d\tau} \prod_\alpha I_{x_\alpha}(\tau)\right], 
\end{equation}
where the $I_n$ are modified Bessel functions, $I_n(\tau) =
\frac{1}{2\pi} \int_0^{2\pi} \rmd k \, \cos(n k) \rme^{\tau \cos k}$.
Choosing $v$ to ensure that $\mathcal{Q}_d(\bx)=1$ for $\bx\in
N(\boldsymbol{0})$ then gives for $\bx\neq \boldsymbol{0}$, 
\begin{equation}
\mathcal{Q}_d(\bx) = 1-\mathcal{P}_d(\bx) = 
\frac{\int_0^\infty \rmd\tau\, \rme^{-d\tau} \prod_\alpha
I_{x_\alpha}(\tau)}
{\int_0^\infty \rmd\tau\, \rme^{-d\tau}  I_0^{d-1}(\tau) I_1(\tau)}. 
  \label{equ:pinf}
\end{equation}
We remark that for $\bx=\boldsymbol{0}$ one finds, by retaining the
$\delta_{\bx,\boldsymbol{0}}$ term, that
$\mathcal{Q}_d(\boldsymbol{0})=1$ as it must be: starting from the
origin, already the first move reaches an absorption site. 
Given the derivation of (\ref{equ:pinf}), it is not too surprising
that the result is similar to that for the standard 
case of absorption at the origin, where one finds the same expression
but with the denominator replaced by $\int \rmd\tau\,\rme^{-d \tau}
I_0^d(\tau)$.

In dimensions $d=1,2$ the original pair of walkers 
always coagulate eventually, regardless of the initial separation 
and thus $\mathcal{P}_d(\bx)=0$. (This can be seen formally from the
fact that both the numerator and denominator integrals in (\ref{equ:pinf}) are
dominated by their divergent tails; these have $\bx$-independent
prefactors, giving $\mathcal{Q}_d(\bx)=1$.)
In $d>2$, on the other hand, 
Equation~(\ref{equ:pinf}) yields a 
nonzero probability $\mathcal{P}_d(\bx)$ that 
the walkers survive indefinitely. The two particular values we need in the main 
text are 
\begin{eqnarray}
  \mathcal{P}_3(2,0,0) & = &  1 - \frac
  {\int_0^\infty \rmd \tau \, \rme^{-3 \tau} I_0^2(\tau) I_2(\tau)}
  {\int_0^\infty \rmd \tau \, \rme^{-3 \tau} I_0^2(\tau) I_1(\tau)} \approx 0.50166,
  \label{equ:p200} \\ 
  \mathcal{P}_3(1,1,0) & = &  1 - \frac
  {\int_0^\infty \rmd \tau \, \rme^{-3 \tau} I_0  (\tau) I_1^2(\tau)}
  {\int_0^\infty \rmd \tau \, \rme^{-3 \tau} I_0^2(\tau) I_1  (\tau)} \approx 0.35872.
  \label{equ:p110}   
\end{eqnarray}
The rather significant difference between the numerical values 
(\ref{equ:p200}) and (\ref{equ:p110}) has a simple reason: when starting at 
initial separation $\bx = (1,1,0)$ there are 12 possible first moves since
each walker has 6 NN sites to move to. Out of these, 4 lead to the
walkers being on NN sites where they coagulate instantaneously, while
for $\bx = (2,0,0)$ only 2 moves produce this outcome. There are similar
differences in subsequent moves which accumulate to the numbers given above.

\end{document}